\documentclass[aps,prd,twocolumn,superscriptaddress,showpacs]{revtex4-1}

\usepackage{graphicx}
\usepackage{dcolumn}
\usepackage{bm}
\usepackage{hyperref}

\usepackage{amsmath}
\usepackage{amssymb}
\usepackage{epstopdf}

\begin{document}


\title{Measurement of Cosmic-ray Muons and Muon-induced Neutrons in the\\Aberdeen Tunnel Underground Laboratory}


\author{S.~C.~Blyth}
\affiliation{Department of Electro-Optical Engineering, National United University, Miao-Li 36063, Taiwan}
\author{Y.~L.~Chan}
\affiliation{Department of Physics, Chinese University of Hong Kong, Hong Kong, China}
\author{X.~C.~Chen}
\affiliation{Department of Physics, Chinese University of Hong Kong, Hong Kong, China}
\author{M.~C.~Chu}
\affiliation{Department of Physics, Chinese University of Hong Kong, Hong Kong, China}
\author{K.~X.~Cui}
\affiliation{Department of Physics, University of Hong Kong, Hong Kong, China}
\author{R.~L.~Hahn}
\affiliation{Chemistry Department, Brookhaven National Laboratory, Upton, NY 11973, USA}
\author{T.~H.~Ho}
\affiliation{Department of Physics, National Taiwan University, Taipei 10617, Taiwan}
\author{Y.~K.~Hor}
\affiliation{Department of Physics, Chinese University of Hong Kong, Hong Kong, China}
\author{Y.~B.~Hsiung}
\affiliation{Department of Physics, National Taiwan University, Taipei 10617, Taiwan}
\author{B.~Z.~Hu}
\affiliation{Institute of Physics, National Chiao Tung University, Hsinchu 300, Taiwan}
\author{K.~K.~Kwan}
\affiliation{Department of Physics, Chinese University of Hong Kong, Hong Kong, China}
\author{M.~W.~Kwok}
\affiliation{Department of Physics, Chinese University of Hong Kong, Hong Kong, China}
\author{T.~Kwok}
\affiliation{Department of Physics, University of Hong Kong, Hong Kong, China}
\author{Y.~P.~Lau}
\affiliation{Department of Physics, University of Hong Kong, Hong Kong, China}
\author{K.~P.~Lee}
\affiliation{Department of Physics, University of Hong Kong, Hong Kong, China}
\author{J.~K.~C.~Leung}
\affiliation{Department of Physics, University of Hong Kong, Hong Kong, China}
\author{K.~Y.~Leung}
\affiliation{Department of Physics, University of Hong Kong, Hong Kong, China}
\author{G.~L.~Lin}
\affiliation{Institute of Physics, National Chiao Tung University, Hsinchu 300, Taiwan}
\author{Y.~C.~Lin}
\affiliation{Department of Physics, Chinese University of Hong Kong, Hong Kong, China}
\author{K.~B.~Luk}
\affiliation{Department of Physics, University of California at Berkeley, Berkeley, CA 94720, USA}
\author{W.~H.~Luk}
\affiliation{Department of Physics, Chinese University of Hong Kong, Hong Kong, China}
\author{H.~Y.~Ngai}
\email[Corresponding author.\\Electronic address (H.~Y.~Ngai): ]{jngai@graduate.hku.hk}
\affiliation{Department of Physics, University of Hong Kong, Hong Kong, China}
\author{W.~K.~Ngai}
\affiliation{Department of Physics, Chinese University of Hong Kong, Hong Kong, China}
\author{S.~Y.~Ngan}
\affiliation{Department of Physics, Chinese University of Hong Kong, Hong Kong, China}
\author{C.~S.~J.~Pun}
\affiliation{Department of Physics, University of Hong Kong, Hong Kong, China}
\author{K.~Shih}
\affiliation{Department of Physics, Chinese University of Hong Kong, Hong Kong, China}
\author{Y.~H.~Tam}
\affiliation{Department of Physics, Chinese University of Hong Kong, Hong Kong, China}
\author{R.~H.~M.~Tsang}
\affiliation{Department of Physics, University of Hong Kong, Hong Kong, China}
\author{C.~H.~Wang}
\affiliation{Department of Electro-Optical Engineering, National United University, Miao-Li 36063, Taiwan}
\author{C.~M.~Wong}
\affiliation{Department of Physics, Chinese University of Hong Kong, Hong Kong, China}
\author{H.~H.~C.~Wong}
\affiliation{Department of Physics, University of Hong Kong, Hong Kong, China}
\author{H.~L.~H.~Wong}
\affiliation{Department of Physics, University of Hong Kong, Hong Kong, China}
\author{K.~K.~Wong}
\affiliation{Department of Physics, Chinese University of Hong Kong, Hong Kong, China}
\author{M.~Yeh}
\affiliation{Chemistry Department, Brookhaven National Laboratory, Upton, NY 11973, USA}

\collaboration{Aberdeen Tunnel Experiment Collaboration}
\noaffiliation

\date{\today}

\begin{abstract}
We have measured the muon flux and production rate of muon-induced neutrons at a depth of 611 m water equivalent. Our apparatus comprises three layers of crossed plastic scintillator hodoscopes for tracking the incident cosmic-ray muons and 760 L of gadolinium-doped liquid scintillator for producing and detecting neutrons. The vertical muon intensity was measured to be $I_{\mu} = (5.7 \pm 0.6) \times 10^{-6}$ cm$^{-2}$s$^{-1}$sr$^{-1}$. The yield of muon-induced neutrons in the liquid scintillator was determined to be $Y_{n} = (1.19 \pm 0.08 \textnormal{(stat)} \pm 0.21 \textnormal{(syst)}) \times 10^{-4}$ neutrons/($\mu\cdot$g$\cdot$cm$^{-2}$). A fit to the recently measured neutron yields at different depths gave a mean muon energy dependence of $\left\langle E_{\mu} \right\rangle^{0.76 \pm 0.03}$ for liquid-scintillator targets.
\end{abstract}


\maketitle

\section{Introduction\label{sec:intro}}

Besides photons and neutrinos, muons are the most abundant secondary cosmic radiation at sea level. The integrated muon flux through a horizontal plane at sea level is about 1 cm$^{-2}$min$^{-1}$ \cite{bib:olive}. High-energy muons can penetrate into underground and generate background to some sensitive experiments, such as dark matter searches, low-energy neutrino oscillation experiments, and neutrinoless double beta-decay experiments. Muons can be easily identified and vetoed, and so they usually do not directly constitute a serious background. However, muons can still affect the experiments in several ways \cite{bib:formaggio}. First, vetoing muons increases the dead time of an experiment, particularly for shallow sites where the muon rates are high. Second, low-energy negative muons can be captured by nuclei, and give rise to neutrons and radioactive isotopes. The effect of stopping muons is also more significant at shallow sites. Third, high-energy muons can induce spallation neutrons and radioisotopes. These spallation neutrons have a very broad spectrum that extends up to several GeV in neutron energy. They can travel a long distance into the detector and are difficult to tag. Neutron scattering and capture within the target can mimic the signal. Therefore, understanding the properties of muons and muon-induced neutrons is important to sensitive underground experiments. There are nine measurements on the production of muon-induced neutrons in an organic liquid scintillator at various depths, ranging from 20 to 5200 m water equivalent (m.w.e.) \cite{bib:hertenberger, bib:bezrukov, bib:boehm, bib:bezrukov, bib:enikeev, bib:abe, bib:persiani, bib:bellini, bib:aglietta-lsd}. In general, such measurements require detailed and experiment-specific Monte Carlo simulations to correct for the neutron contribution from rock as well as the metallic components of the detectors, but this is difficult, especially for the older experiments like those in Refs.~\cite{bib:bezrukov, bib:enikeev, bib:aglietta-lsd}. Recent measurements \cite{bib:abe, bib:persiani, bib:bellini} are all byproducts of advanced neutrino experiments, which are carried out at great depths ($>$2500 m.w.e.). The Aberdeen Tunnel experiment is dedicated to measure the muon flux and production rate of muon-induced neutrons in an organic liquid scintillator at a relatively shallow depth.

\begin{figure}
	\centering
	\includegraphics[width=\linewidth]{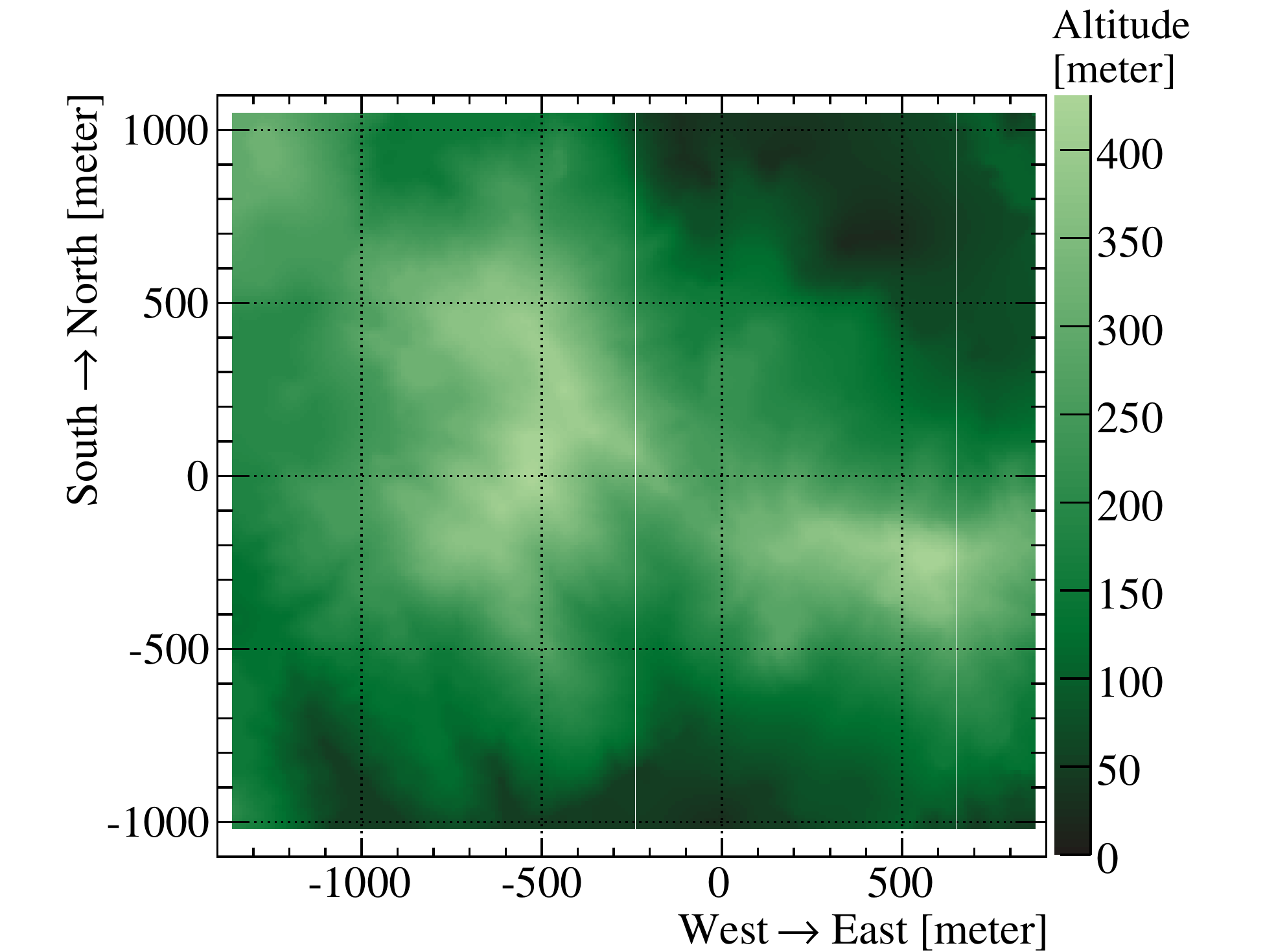}
	\caption{Contour map of the mountains above the Aberdeen Tunnel laboratory, which is located at (0, 0).}
	\label{fig:mount-profile}
\end{figure}

The Aberdeen Tunnel laboratory is located inside the middle cross-passage of the Aberdeen Tunnel, which is a 1.9-km-long two-tube vehicle tunnel, in Hong Kong. It is beneath the saddle-shaped landscape between Mount Nicholson (on the east, 430 m tall) and Mount Cameron (on the west, 439 m tall). A contour map of the two mountains is shown in Fig.~\ref{fig:mount-profile}. The laboratory is 22 m above sea level at $22.23^{\circ}$N and $114.6^{\circ}$E and has an overburden of approximately 235 m of rocks, or 611 m.w.e. Since Mount Nicholson and Mount Cameron are both over 400 m in height, the slant overburden in the southeast (due to Mount Nicholson) and west (due to Mount Cameron) directions can go up to 700--800 m.w.e.

\newpage

In this paper, we present the results of a study of cosmic-ray muons and their induced neutrons in the Aberdeen Tunnel laboratory. The following subsections briefly describe the apparatus and the data acquisition system. A detailed description of the experimental site and the setup can be found in Ref.~\cite{bib:abt-nim}. Section \ref{sec:analysis} outlines the algorithms for event reconstruction and describes the scheme for event selection. In Sec.~\ref{sec:eff-uncert}, we discuss the detection efficiencies and the corresponding systematic uncertainties. The results of the muon flux and the muon-induced neutron yield are presented in Sec.~\ref{sec:result}. Finally, a summary of our main findings is given in Sec.~\ref{sec:conclude}.

\subsection{Apparatus\label{sec:app}}

\begin{figure}
	\centering
	\includegraphics[width=\linewidth]{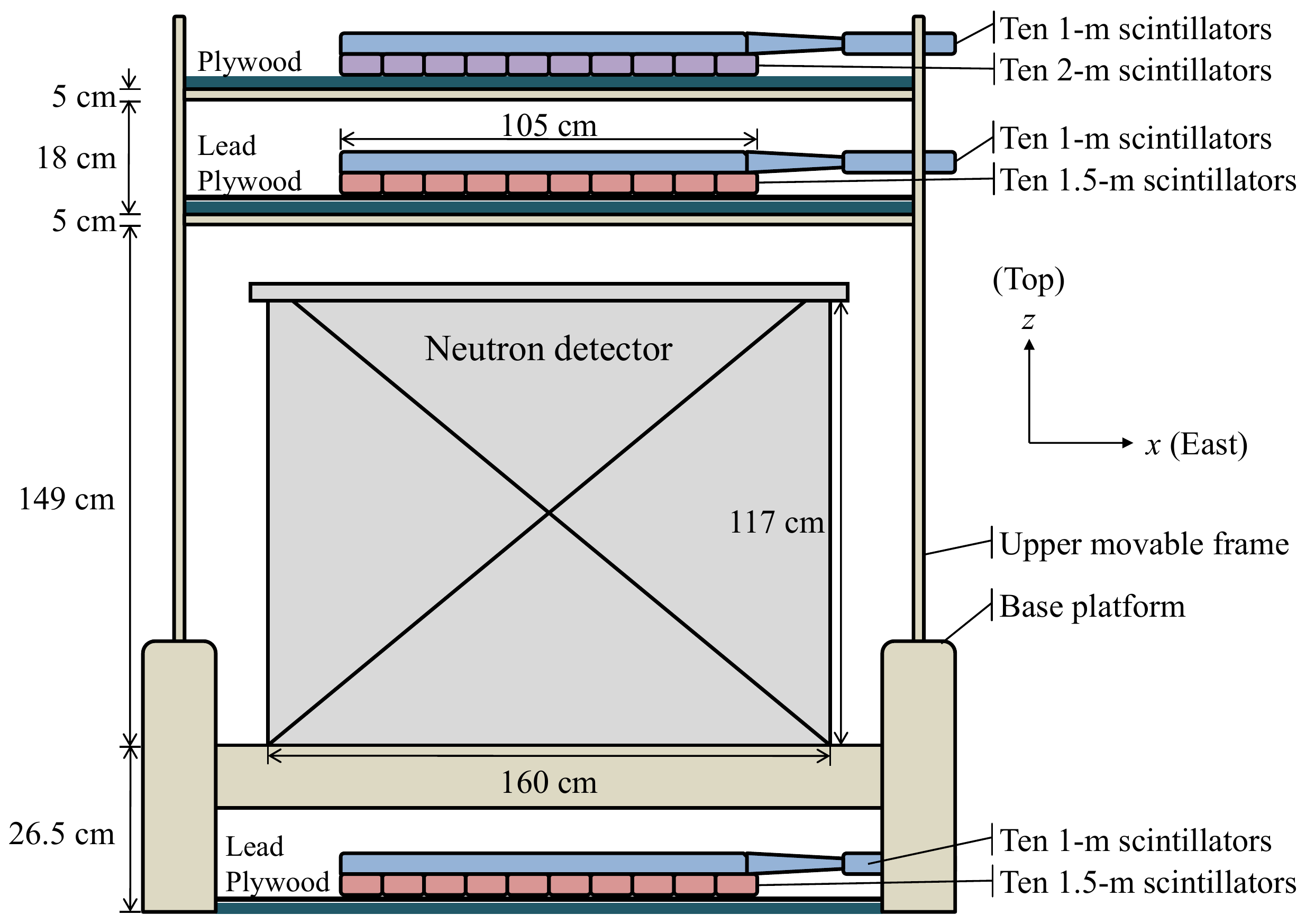}
	\caption{Schematic drawing of the apparatus. The neutron detector is sandwiched in between the hodoscopes of the muon tracker. We used a Cartesian coordinate system and defined positive $x$ to be due east and positive $y$ to be due north. The origin of the coordinate system was set at the center of the neutron detector.}
	\label{fig:app-front-view}
\end{figure}

The apparatus consists of a muon tracker (MT) and a neutron detector (ND) as shown in Fig.~\ref{fig:app-front-view}. The MT tracks the directions and positions of the incoming cosmic-ray muons and provides the triggers for detecting muon-induced neutrons in the ND. The central vertical axis of the ND was aligned with the center of each hodoscope plane of the MT. The relative position of the two detectors was aligned to be better than 2 mm. An outline of each detector is given in the following subsections.

\subsubsection{Muon tracker\label{sec:app-mt}}

The MT consists of three layers. Each layer is made up of two overlapping planes of orthogonally arranged plastic scintillator hodoscopes for determining the coordinates of a muon. In each layer, one plane consists of 10 1-m-long hodoscopes, while the other plane consists of 10 either 2-m-long (in the top layer) or 1.5-m-long (in the middle and the bottom layers) hodoscopes. Each hodoscope, except the 2-m-long ones, is formed by a rectangular plastic scintillator, a photomultiplier tube (PMT), and a trapezoidal Lucite light guide between the scintillator and the PMT, while each 2 m hodoscope has a PMT and a light guide on both ends of the plastic scintillator. The scintillators are covered by reflective aluminium foils to increase the collection efficiency of scintillation light. The hodoscopes are then wrapped in opaque black plastic sheets to prevent light leakage. The top and the middle layers are put above the neutron detector. The bottom layer rests on the floor, which is covered with sheets of lead and plywood. The three layers of hodoscopes and the neutron detector are aligned vertically.

\subsubsection{Neutron detector\label{sec:app-nd}}

\begin{figure}
	\centering
	\includegraphics[width=\linewidth]{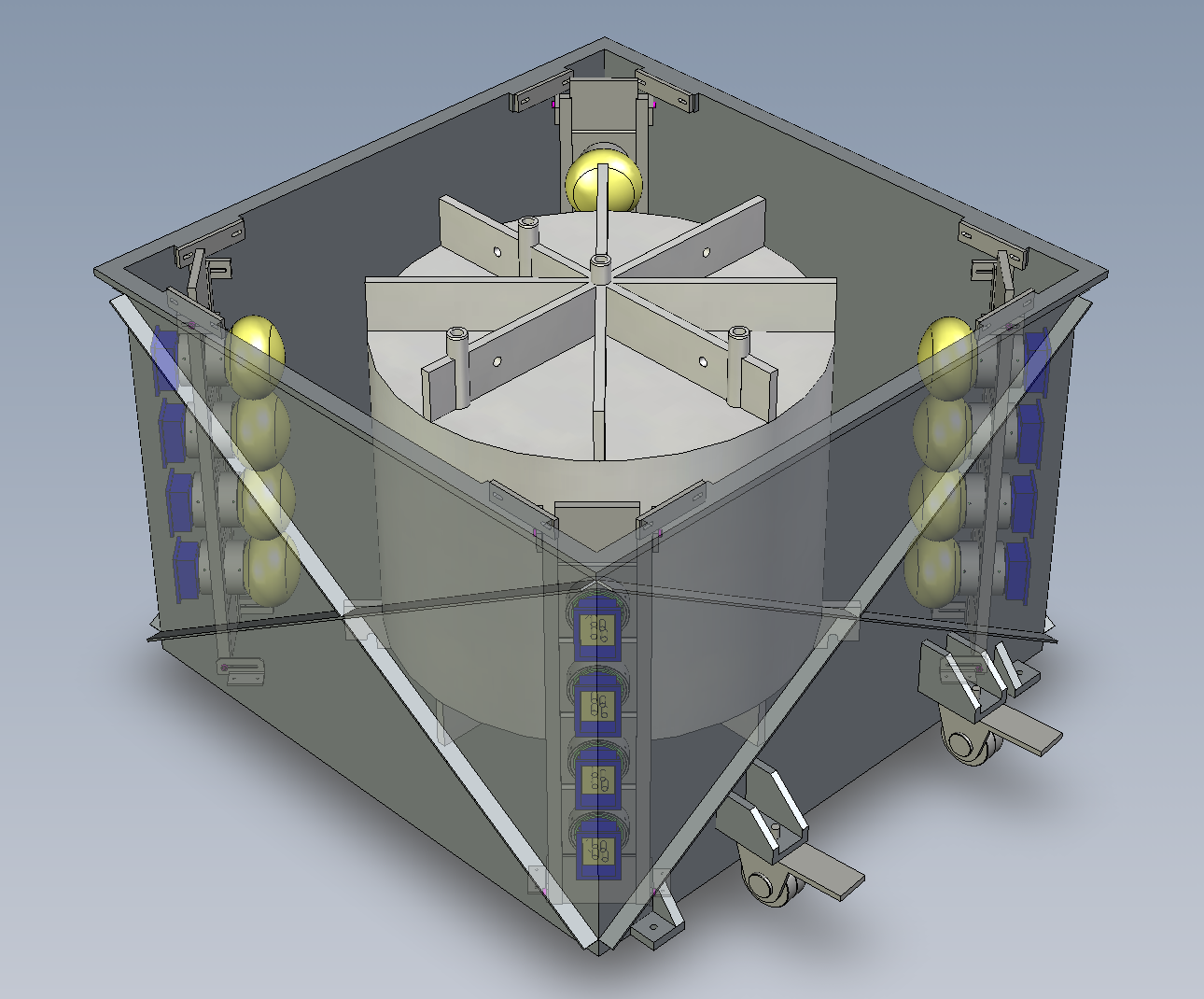}
	\caption{Schematic drawing of the ND. The top cover and all the reflectors are not shown for simplicity. Sixteen PMTs are located at the four corners of a stainless steel tank facing the acrylic vessel.}
	\label{fig:nd-structure}
\end{figure}

The ND is a calorimeter. It employs a two-zone design. The outer zone contains 1900 L (1.63 tonne) of mineral oil, used as a buffer to attenuate gamma rays from outside to enter the target volume and to suppress ambient slow-neutron backgrounds. The inner zone contains 760 L (0.65 tonne) of 0.06\% gadolinium-doped linear-alkyl-benzene-based liquid scintillator (Gd-LS), which is the target for neutron production and for detecting neutrons. The two zones are separated by a cylindrical acrylic vessel with an inner diameter of 110 cm. The thickness of the vessel is 1 cm for the vertical surface and 1.5 cm for both the top and bottom plates. Neutron capture on gadolinium gives rise to multiple gamma rays with a total energy of about 8 MeV, which is significantly higher than the energy of background gamma rays. Scintillation photons created by the gamma rays are detected with 16 Hamamatsu R1408 20 cm PMTs, which are located at the four corners of the ND as shown in Fig.~\ref{fig:nd-structure}. To increase the total number of detected photons, hence improving the energy resolution, a total of two circular specular reflectors of 140 cm diameter each are put above and beneath the acrylic vessel, respectively. Four diffusive-reflector panels are mounted on the inner walls of the ND to further improve the energy resolution and the uniformity of energy response across the target volume. There are three calibration ports on the top of the ND for deploying calibration sources, namely, the center port, the north port (25 cm away from the center), and the south port (45 cm away from the center).

\subsection{Data acquisition and triggers\label{sec:daq}}

\begin{figure*}
	\centering
	\includegraphics[width=15cm]{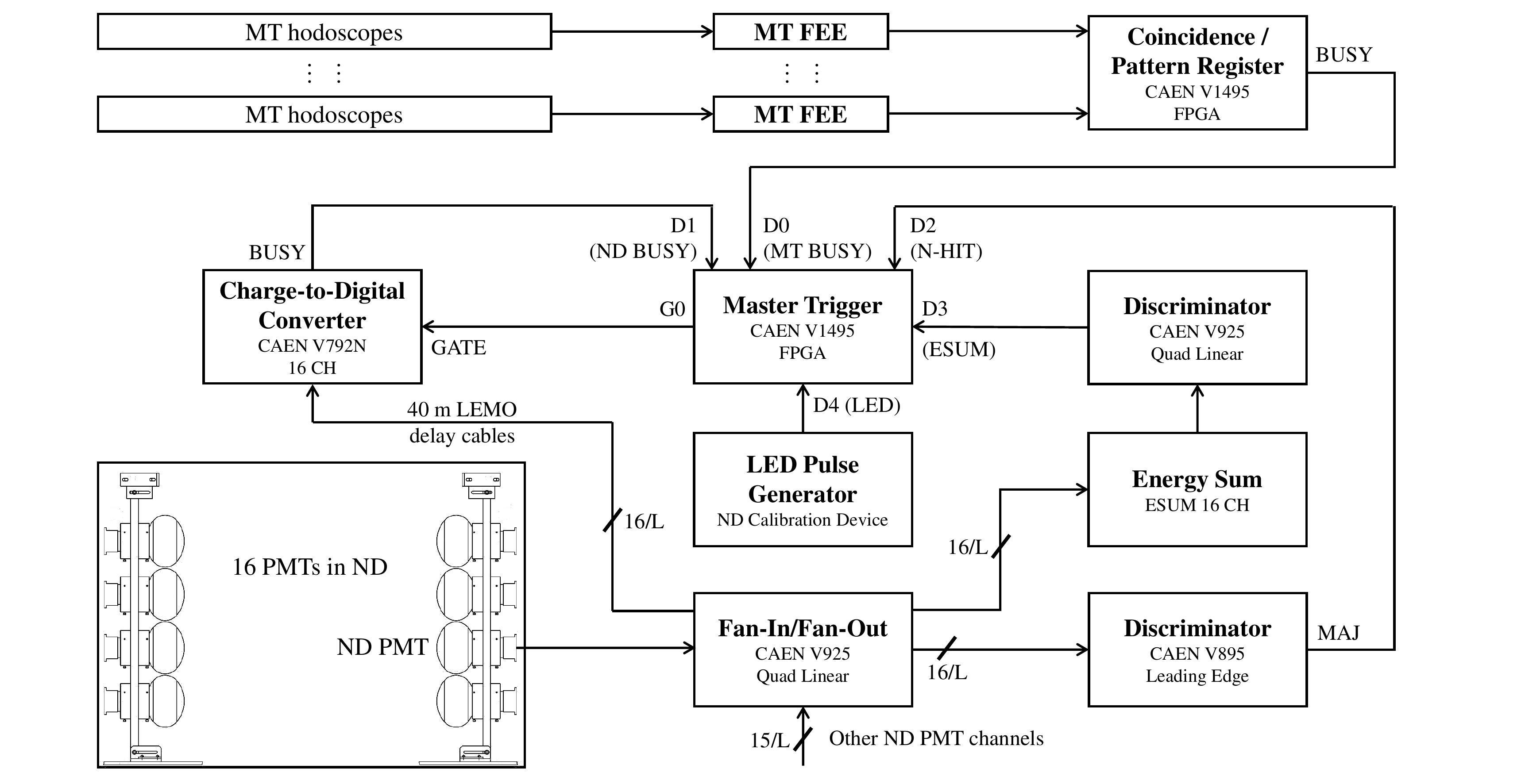}
	\caption{Block diagram of the data acquisition system.}
	\label{fig:daq-block-diagram}
\end{figure*}

The data acquisition (DAQ) system was set up as shown in Fig.~\ref{fig:daq-block-diagram}. The PMT signals from the MT are digitized by the front-end electronics (FEE). A coincidence-and-pattern-register module handles the signals from every MT FEE according to a multiplicity trigger condition. To reconstruct a muon track, at least two coordinates are required in each of the orthogonal directions. Therefore, the double 2-out-of-3 multiplicity (also known as ``2/3-X and 2/3-Y'') trigger is the minimal trigger condition. The output of the DAQ for the MT is a binary map (a ``hit pattern'') showing which hodoscopes have hits in coincidence.

For the ND, each PMT signal is duplicated into three copies by a linear fan-in/fan-out module. One copy goes directly into a charge-to-digital converter (QDC) for charge measurement. Another copy goes into a leading-edge discriminator. When the number of channels with a waveform exceeding the preset threshold is greater than or equal to a designated number, a logic signal is generated at the majority (MAJ) output. This majority threshold determines the multiplicity (N-HIT) trigger of the ND. The remaining copy goes into an analog energy-sum (ESUM) module to integrate the charges from all the 16 PMTs. The output of the ESUM module goes into a discriminator which determines the energy threshold of the ESUM trigger. The logic signals of the N-HIT trigger, the ESUM trigger, and a light-emitting-diode (LED) trigger from the ND calibration device are passed to a Master Trigger Board (MTB) for the final trigger decision. The MTB is realized by using a field-programmable gate array running at 100 MHz. It can also generate an optional periodic trigger for the ND to monitor the QDC pedestals. To reduce the potential energy dependence of the trigger efficiency due to variation of the PMT gain, the N-HIT trigger is used as the primary trigger. The N-HIT threshold can be set from 1 out of 16 to 16 out of 16. A prescaled ESUM trigger with threshold of about 0.5 MeV is used to monitor background events.

Busy signals from the DAQ subsystems of the MT and the ND are generated during event building, charge conversion, or when the event buffer is full. Thus, a busy signal also represents an accepted trigger. The MTB time stamps the falling edge and the rising edge of the busy signals with 10 ns time resolution and records the corresponding event type (MT or ND) and, for ND events, also records the trigger type (N-HIT, ESUM, LED, or periodic). Events can be correlated in offline analysis using the time stamps to search for muon-induced neutrons. The widths of the busy signals can be used to determine the dead time of the experiment precisely.

The front-end electronics are connected to a front-end computer via a CAEN V1718 VME-USB2.0 interface. The front-end computer is also connected to a CAEN SY1527LC high-voltage system, an environmental temperature and humidity sensor, a temperature sensor for the ND, and a motorized calibration device for the ND. Run control is done with an open-source data acquisition software called MIDAS \cite{bib:ritt-midas}. A back-end computer running the MIDAS server and data logger is linked to the front-end computer through a 100BASE-TX Ethernet. Events are stored using the MIDAS format and compressed with GNU-zip \cite{bib:gnu-zip}. The compression reduces the average size of an event by approximately 60\% to about 40 bytes. Data files are written to a 500 GB hard disk in the back-end computer and are regularly transferred from the laboratory to a 6 TB RAID-5 disk-array (i.e., redundant array of independent disks with distributed parity) in the University of Hong Kong for processing.

\subsection{Calibration of the neutron detector\label{sec:calib}}

Energy calibration is performed regularly by deploying radioactive calibration sources ($^{137}$Cs, $^{60}$Co, and $^{241}$Am-Be) in the center of the ND. The $^{137}$Cs provides 0.66 MeV gamma rays, and the $^{60}$Co emits gamma rays of 1.17 and 1.33 MeV. The $^{241}$Am-Be is used as a neutron source. The process of neutron capture and the efficiency for detecting neutrons are studied by deploying the $^{241}$Am-Be at various positions inside the ND. The source produces neutrons predominantly through the reaction 
\begin{equation}
	^{9}\textnormal{Be} + \alpha \to ^{13}\textnormal{C}^{*} \to ^{12}\textnormal{C} + n + \gamma \textnormal{(4.4 MeV)} .
	\label{eq:ambe-reaction-w-gamma}
\end{equation}
The 4.4 MeV gamma ray and the subsequent neutron capture form a distinct signature that can be selected by a delayed coincidence technique. The 4.4 MeV gamma rays are usually detected together with the proton-recoil signals of roughly 1 MeV, which are generated in the thermalization process of the neutrons. This forms the prompt signal, and the gamma rays emitted from the neutron-capture process give rise to a delayed signal. The neutrons from the $^{241}$Am-Be source were selected by requiring the reconstructed energy of the prompt signals to be between 4.4 and 6.4 MeV and a temporal separation between the prompt and delayed signals of less than 200 $\mu$s. Neutrons in the Gd-LS give rise to two prominent gamma-ray energy peaks at 2.2 MeV and around 8 MeV due to the capture of neutrons on hydrogen and gadolinium, respectively. The peak around 8 MeV is a result of two gadolinium isotopes with similar energies of the emitted gamma rays, namely, $^{155}$Gd and $^{157}$Gd with total gamma-ray energies of 8.54 and 7.94 MeV, respectively.

Relative gains of the PMTs are determined by dividing the measured QDC responses due to a calibration source at the center of the ND by the corresponding expected values. The expected values are calculated with an optical model of the ND which considers the reflection of the reflectors and the attenuation of the liquids. This method of determining relative gains has been cross-checked by a planar LED light source which emits light uniformly in a horizontal plane. Both methods gave consistent results for PMTs in the same horizontal plane.

\section{Data analysis\label{sec:analysis}}

\subsection{Raw data\label{sec:raw-data}}

\begin{figure}
	\centering
	\includegraphics[width=\linewidth]{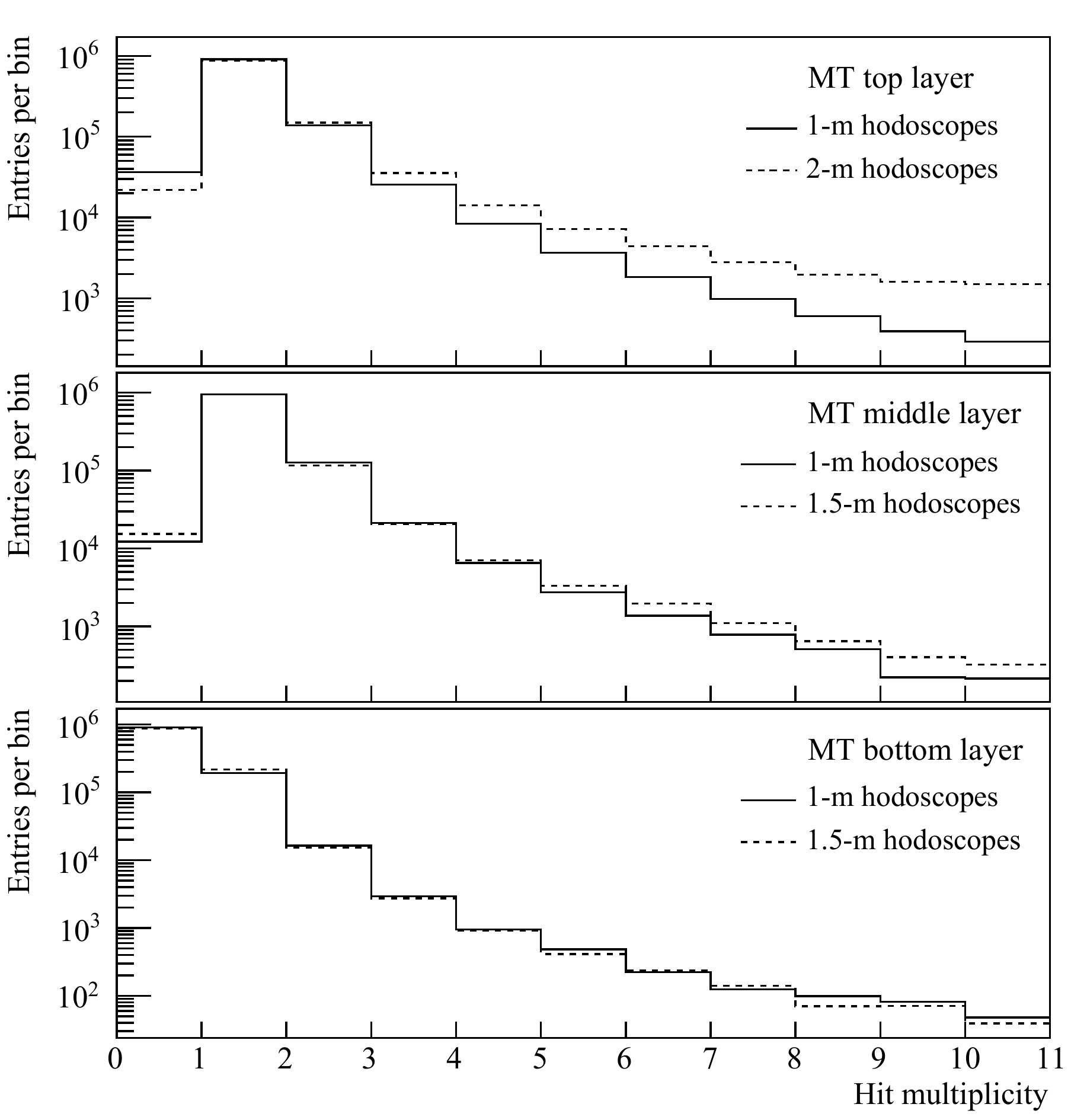}
	\caption{Raw hit multiplicity of each MT hodoscope plane.}
	\label{fig:mt-hit-multiplicity}
\end{figure}

One million MT events collected in 190 days of experiment live time during the year of 2012 were analyzed. The trigger rate of the MT was about 0.065 Hz. Figure \ref{fig:mt-hit-multiplicity} shows the raw hit-multiplicity of each hodoscope plane. The higher rate of the 2-m-long hodoscopes for hit multiplicity $>$ 3 could not be explained by the coincidence of noise hits. Thus, it was believed that the 2 m hodoscopes, each of which have PMTs on both ends, were more sensitive to the secondary particles induced by muons. The higher rate of zero hits in the bottom layer was due to the limited acceptance of the bottom layer under the trigger condition of 2/3-X and 2/3-Y, in which about 80\% of the triggered events involved only the top and the middle layers. The fraction of clean hits (i.e., multiplicity = 1) over all nonzero hits in each layer was 82\%, 86\%, and 91\% for the top, middle, and bottom layers, respectively. This increasing trend of clean hits was due to the attenuation of the secondary particles that were generated from the rocks by the incident muons.

\begin{figure}
	\centering
	\includegraphics[width=\linewidth]{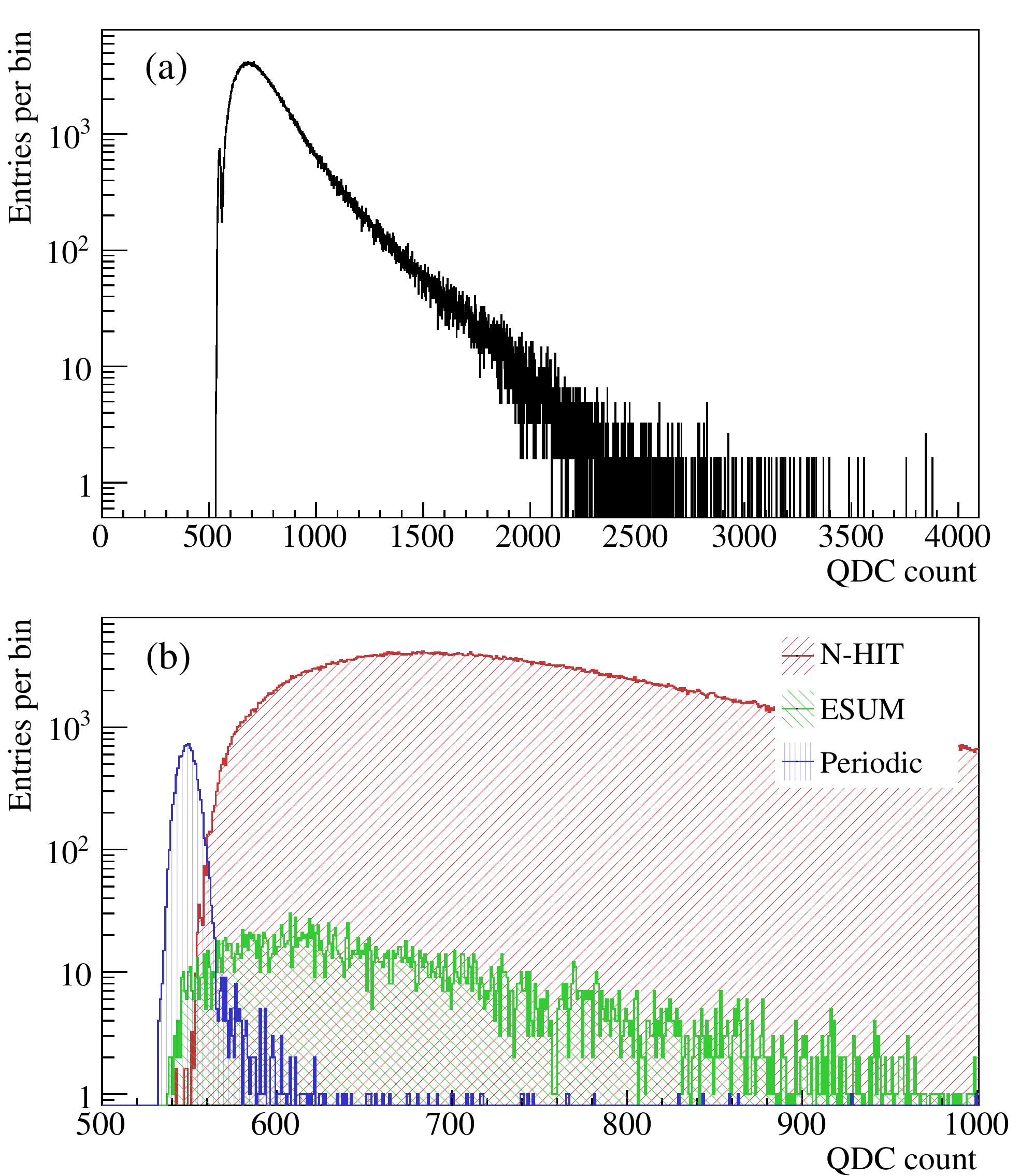}
	\caption{(a) A typical charge histogram from the QDC output of a ND PMT regardless of the trigger type. (b) Charge histograms of different trigger types with the QDC count limited to 500--1000 to show the details at low energies.}
	\label{fig:nd-charge}
\end{figure}

The same set of data used for measuring the muon flux was further analyzed for studying the muon-induced neutrons. The N-HIT trigger rate of the ND, with $N$ being set to 16, was about 6 kHz. The ESUM trigger was prescaled by a factor of 1000 to a rate of about 17 Hz. The periodic trigger was set to 50 Hz. During data taking, data was reduced by a factor of 150 by only keeping ND events within 100 ms after the preceding MT trigger. In the offline analysis, data was further reduced by a factor of 30 by only keeping ND events within 3 ms following the preceding MT event. Examples of charge histograms of the three trigger types after the data reduction are shown in Fig.~\ref{fig:nd-charge}. The figures were drawn from about 13 days of experimental data, which contains roughly one million ND events.

\subsection{Muon track reconstruction\label{sec:recon-mt}}

The muon tracks provide information on their angular distributions and facilitate the analysis of muon-induced neutrons by giving the path lengths in the target volume. The goal of track reconstruction is to find a muon trajectory where its expected hit pattern most closely resembles the observed hit pattern of the MT. A three-dimensional track reconstruction is done by decomposing the track into two two-dimensional projections on the x-z and the y-z planes, respectively. The two-dimensional tracks are reconstructed by linear regression using the coordinates of the fired hodoscopes. Occasionally, a muon can fire two adjacent hodoscopes in the same layer. Therefore, the first stage of reconstruction is to identify and combine those adjacent fired hodoscopes into a cluster. Every cluster is treated as a single hodoscope with extended width. Coordinates used in the linear regressions are sampled randomly within the width of the clusters to avoid favouring some particular track directions. In each event, up to ten coordinates are sampled from a cluster, and multiple straight lines are reconstructed from all possible combinations of these coordinates. Then, the reconstructed lines in both the x-z and y-z planes are merged to form three-dimensional tracks using all possible combinations. Finally, every track candidate is evaluated and assigned with a ``fitness'' value, which represents its likelihood to induce the observed hit pattern of the MT.

\begin{figure}
	\centering
	\includegraphics[width=\linewidth]{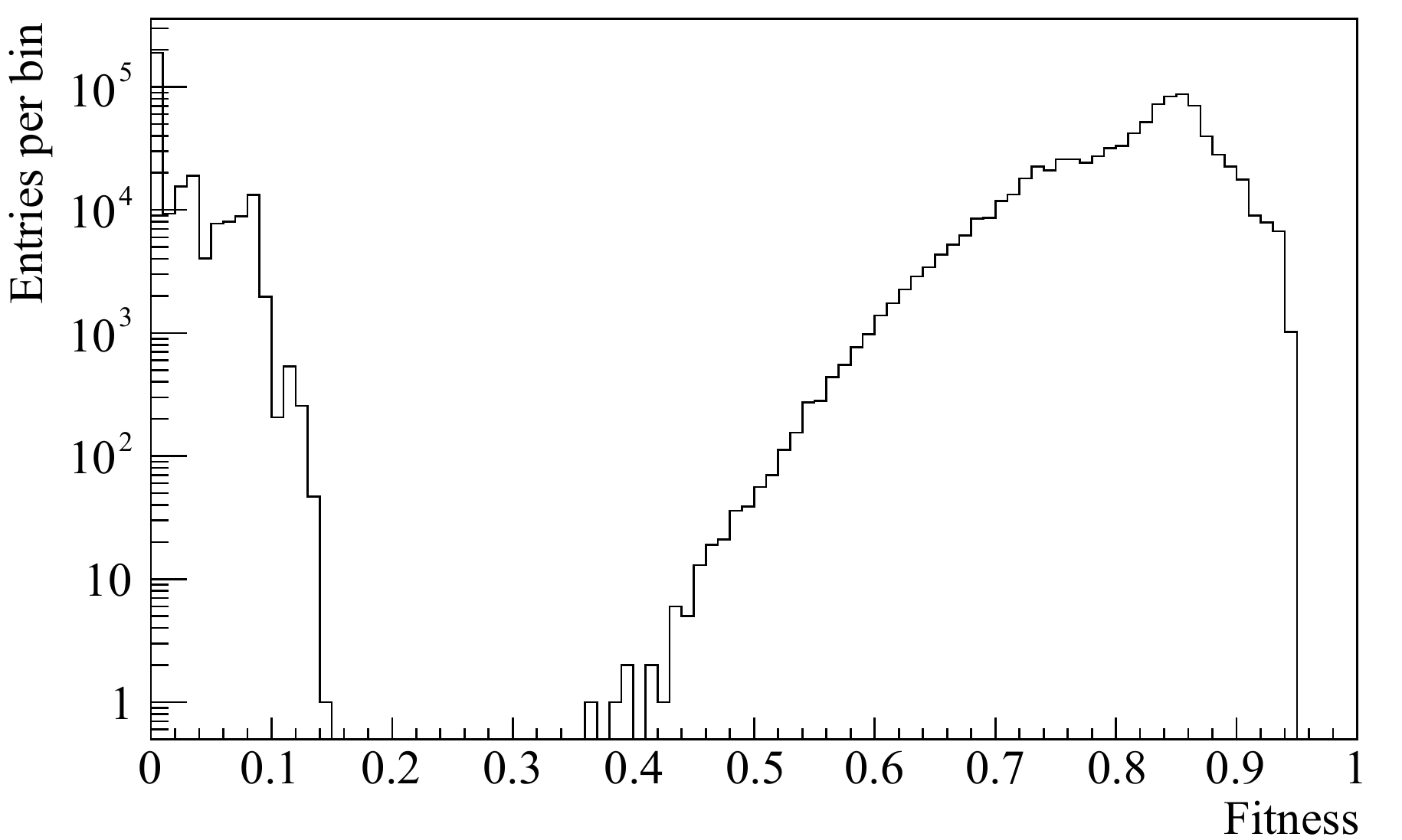}
	\caption{Fitness distribution of the reconstructed muon tracks under the trigger condition of 2/3-X and 2/3-Y.}
	\label{fig:mu-fitness}
\end{figure}

The first step of fitness evaluation is to simulate the hit pattern due to the track candidate. It is done by simple geometrical consideration to find out all the hodoscopes that are intercepted by a given track. Suppose the actual hit pattern of the MT and the expected hit pattern due to the track candidate are represented by binary maps $A$ and $E$, respectively. A binary state of 1 indicates that the hodoscope is fired due to whatever reasons, while a binary state of 0 indicates that the hodoscope is not fired. The fitness $f$ of a track candidate $T$ is defined as
\begin{equation}
	f(T) = \prod_{i=1}^{N_{H}} P(A_{i}, E_{i}(T)) ,
\end{equation}
where $N_{H}$ is the total number of hodoscopes in the MT. $P$ is a conditional function defined as
\begin{equation}
	P(A_{i}, E_{i}) = \left\{ \begin{array}{l l}
		\varepsilon_{e,i} &\quad \textnormal{if } A_{i} = 1 \textnormal{ and } E_{i} = 1 \\
		1 - \varepsilon_{e,i} &\quad \textnormal{if } A_{i} = 0 \textnormal{ and } E_{i} = 1 \\
		1 - \varepsilon_{n,i} &\quad \textnormal{if } A_{i} = 0 \textnormal{ and } E_{i} = 0 \\
		\varepsilon_{n,i} &\quad \textnormal{if } A_{i} = 1 \textnormal{ and } E_{i} = 0 \\
	\end{array} \right. ,
\end{equation}
where $\varepsilon_{e,i}$ and $\varepsilon_{n,i}$ are the efficiency and the noise level of hodoscope $i$, respectively. The noise level is defined as the probability of seeing any noise in coincidence with the trigger. It was estimated from the singles rate of the hodoscope and was found to be less than 0.06\% for every hodoscope. In the analysis, all $\varepsilon_{n,i}$ were set to zero. Thus, a track candidate with any number of outlying hits will have a fitness value equal to zero. On the other hand, a well-defined track candidate without any ambiguity on the hits will have a fitness value greater than zero. The track candidate with the highest fitness value is selected as the solution. The fitness distribution of the best candidate in each event obtained from the muon-flux measurement is shown in Fig~\ref{fig:mu-fitness}. Those tracks with fitness values between 0 and 0.2 were due to one or more null hits in the observed hit pattern because of the finite efficiency of the hodoscopes.

It is convenient to express a reconstructed track with the zenith angle $\theta$ and azimuth angle $\phi$, together with the coordinates $(x,y)$ where the track intercepts the floor. We defined $\phi = 0$ to be due east. The resolution of $\theta$ of the reconstructed tracks under the trigger condition of 2/3-X and 2/3-Y and selected with fitness $>$ 0 was 6.8$^{\circ}$. The resolution of $\phi$ strongly depended on $\theta$. Its value under the same condition was 50$^{\circ}$, 30$^{\circ}$, 15$^{\circ}$, and 7.1$^{\circ}$ for $\theta < 20^{\circ}$, $20^{\circ} < \theta < 40^{\circ}$, $40^{\circ} < \theta < 60^{\circ}$, and $\theta > 60^{\circ}$, respectively. The position resolution is only relevant to the measurement of muon-induced neutrons. Its value for the muons that were selected for the measurement was about 5 cm.

\subsection{Energy reconstruction\label{sec:recon-nd}}

Energy deposited by a particle in the target gives rise to a burst of scintillation photons. Some of the photons are eventually detected by the PMTs inside the ND as amplified charge signals. Visible energy can be estimated by summing over the charges outputted by the PMTs. In practice, the charges are digitized by the QDC. Channel-to-channel differences in the PMT gain and QDC sensitivity are taken into account to reduce the energy nonuniformity across the target volume.

The reconstructed energy $E_{rec}$ calculated with this total-charge method can be expressed as
\begin{equation}
	E_{rec} = \frac{1}{\eta_{E}} \sum_{i=1}^{N_{ch}} \frac{\left(q_{i}-q_{0,i}\right)}{\eta_{G,i}} ,
\end{equation}
where $N_{ch}$ is the number of available QDC channels; $q_{i}$ and $q_{0,i}$ are the output count and the pedestal of QDC channel $i$, respectively; $\eta_{G,i}$ is the overall relative gain of channel $i$; and $\eta_{E}$ is the energy calibration constant, which is defined as the number of QDC counts per unit MeV of energy deposition.

Energy resolution was studied by deploying the default calibration sources ($^{137}$Cs, $^{60}$Co, and $^{241}$Am-Be) as well as $^{54}$Mn and $^{22}$Na to the center of the ND. The prominent gamma-ray energy peaks due to each of the sources, except the $^{241}$Am-Be where only its neutrons were used, were fitted with a Gaussian function to determine the energy resolutions. The peak at 2.2 MeV due to the capture of neutrons on hydrogen was also fitted with a Gaussian function. The 8 MeV peak due to the capture of neutrons on the two gadolinium isotopes was fitted with two Gaussian functions simultaneously. The relative intensity of the gadolinium-isotope peaks ($I_{Gd-155}/I_{Gd-157} = 0.227$) was constrained in the fitting by the abundance and the thermal neutron-capture cross section of the two isotopes. The energy resolution as a function of the reconstructed energy was evaluated to be $\sigma_{E}/E_{rec} = (0.15 \pm 0.03)/\sqrt{E_{rec}}$, with $E_{rec}$ in MeV.

\subsection{Event selection\label{sec:select}}

In the muon-flux measurement, successfully reconstructed muon tracks with fitness values greater than zero were selected. That means the muon events were clean and the reconstructed tracks were well defined. In the measurement of muon-induced neutrons, candidate events were selected by a delayed coincidence technique. The prompt signals were the muon events obtained with the MT, while the delayed signals were the neutron-capture events from the ND. A few more event selection criteria were imposed on the prompt and the delayed signals to reduce the uncertainties in the measurement.

The prompt signals were selected based on the hit topology of the MT. All the 1-m-long hodoscope planes were required to have hits in coincidence to select muons that could pass through the ND. An event was discarded if, in each of the four hodoscope planes in the top and the middle layers, there was more than one cluster or any clusters having more than two hodoscopes. This clean-hit requirement was not applied to the bottom layer because we had to allow for the generation of showers inside the ND. Furthermore, a requirement on the temporal separation between muons was applied to exclude muon events that were present closer than 3 ms to reduce contamination from preceding muons. The delayed signals were selected if the reconstructed energy was greater than 4.6 MeV. This requirement eliminated all the single and most of the two-fold background gamma rays from natural radioactive nuclei, in particular, the 2.6 MeV gamma ray from $^{208}$Tl of the $^{232}$Th series. The time separation between a pair of prompt and delayed signals was required to be within a time window of 10--210 $\mu$s. The first 10 $\mu$s were excluded because the ND had a dead time of a few microseconds after the passage of a muon. Both the energy and the time requirements were optimized simultaneously to minimize the statistical uncertainty in the measured number of muon-induced neutrons. Contribution of background events, such as the ambient fast neutrons from $(\alpha,n)$ reactions and the chance coincidence of background gamma rays, was estimated using a time window between 800 and 1600 $\mu$s after the preceding prompt signal and the same energy requirement for selecting the delayed signal.

\section{Efficiency and systematic uncertainty\label{sec:eff-uncert}}

\subsection{Muon detection efficiency\label{sec:eff-mt}}

Efficiencies of the hodoscopes were measured in a surface laboratory using cosmic-ray muons. The hodoscope being measured and several lead blocks of 5 cm thick were sandwiched in between two other hodoscopes; the signals of these three hodoscopes formed a three-fold coincidence. The coincidence in signals of the two reference hodoscopes, forming a two-fold trigger, indicated the passage of a muon through the middle one. At a discriminator threshold of -50 mV, the efficiency of the sandwiched hodoscope, defined as the ratio of the rate of the three-fold coincidence to the rate of the two-fold coincidence, was measured at different high voltages supplied to the sandwiched hodoscope to determine the efficiency plateau. The plateaued efficiency along the length of the hodoscopes was uniform, with an average efficiency above 95\% for most of the hodoscopes. The efficiency of the MT FEE and the coincidence-and-pattern-register module was measured with rectangular pulses generated from a signal generator and was determined to be close to 100\%.

\begin{figure}
	\centering
	\includegraphics[width=\linewidth]{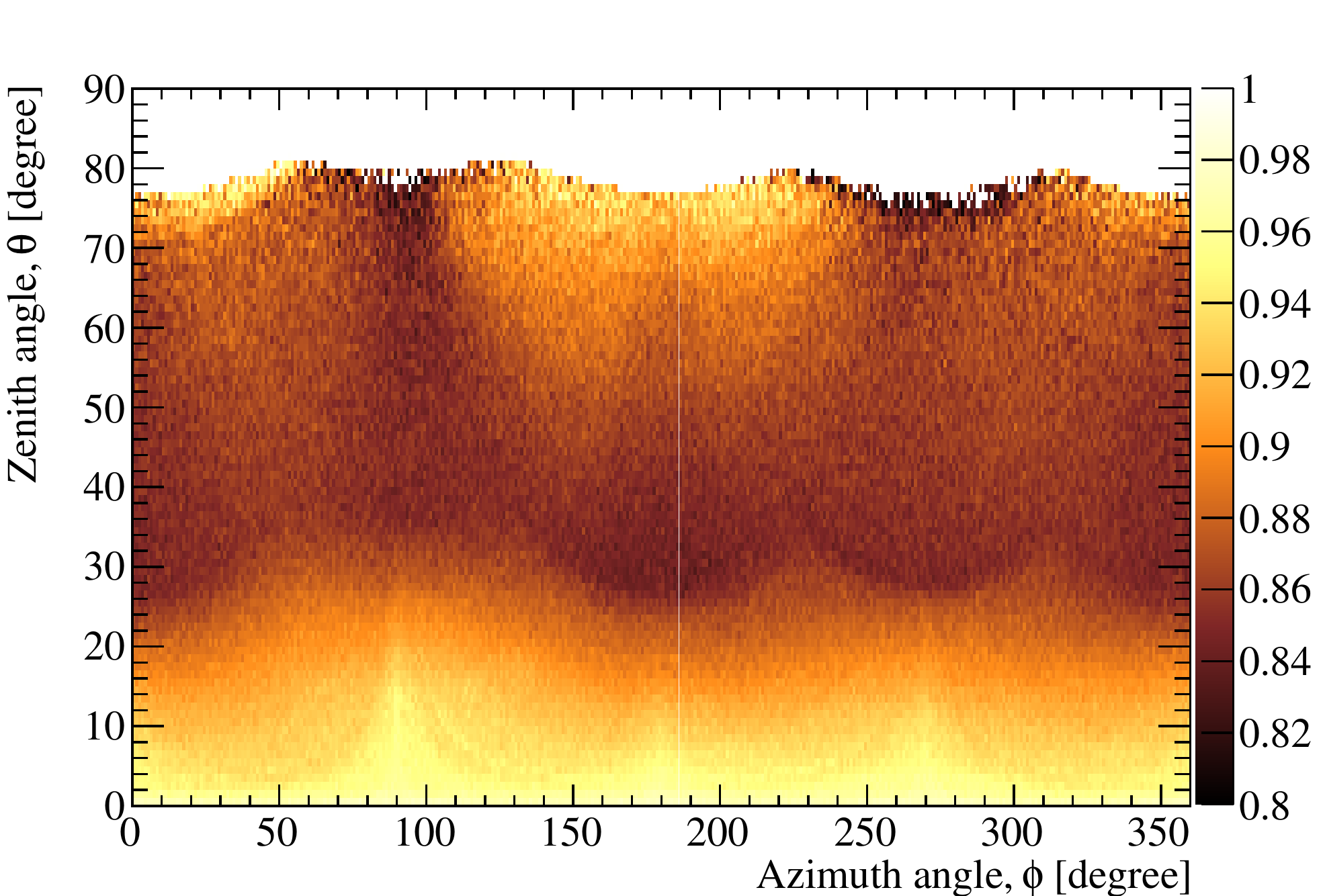}
	\caption{Calculated efficiency of the MT with all the hodoscopes intact and under the trigger condition of 2/3-X and 2/3-Y. The maximum acceptance angle is about 80$^{\circ}$ in zenith. The four tips correspond to the four corners of the MT.}
	\label{fig:mt-efficiency}
\end{figure}

The efficiency of detecting a muon depends on the position and direction where the muon transverses the MT. In the measurement of muon flux, the positional dependence was integrated, and we only considered the angular dependence of the efficiency. The efficiency of the MT, $\varepsilon_{MT}(\theta,\phi)$, at a muon incident angle of $(\theta,\phi)$ was evaluated using Monte Carlo simulation. Muons were generated uniformly from a 3.5 $\times$ 3.5 m plane and perpendicular to the plane. The plane was large enough to cover the entire MT. One muon was generated in each event, and all the hodoscopes that were intercepted by the muon were found. The trigger of the MT was simulated using the acceptance/rejection method with the product of the measured efficiency of the intercepted hodoscopes being the acceptance probability. The number of muons that triggered the MT was counted. The count was then divided by the number obtained by assuming 100\% efficiency in the simulation and counting process to yield the efficiency. The angular dependence of the efficiency under the trigger condition of 2/3-X and 2/3-Y is shown in Fig.~\ref{fig:mt-efficiency}. The MT under this condition had an angular acceptance of about 1.6$\pi$. The efficiency was high, being above 90\%, for small zenith angles because a muon could transverse through all the six planes of hodoscopes. The efficiency was smaller for zenith angles greater than about 30$^{\circ}$ due to the fact that a muon could only hit at most four planes at a time. The efficiency was high again for zenith angles greater than about 70$^{\circ}$ as a muon could transverse multiple hodoscopes in the same plane. The solid-angle weighted average efficiency of the MT was about 87\%. The efficiency of the fitness requirement, $\varepsilon_{Fit}$, under the same condition was directly measured to be about 83\% from the fitness distribution (Fig.~\ref{fig:mu-fitness}).

During the operation of the experiment, some of the hodoscope channels were found to be unstable or dead, resulting in a changing muon detection efficiency over time. These bad channels were masked in the offline analysis according to the unstable or dead periods to make sure that they did not contribute to the measurement. The same set of masks was also applied to the mentioned trigger simulation to derive the corresponding MT efficiency as a function of time. The total systematic uncertainty in the muon-flux measurement was about 10\% mainly due to the uncertainty in the efficiency of the hodoscopes.

\subsection{Neutron detection efficiency\label{sec:eff-nd}}

In the measurement of muon-induced neutrons, our interest is on the neutrons being produced in the Gd-LS. There are three main reasons for the inefficiency of detecting neutrons that are generated inside the target of the ND: 
\begin{itemize}
	\item Neutrons drift out from the target volume without being captured.
	\item Neutron captures occur within the dead time of the DAQ system or outside of the event-selection time windows.
	\item The reconstructed energies of the neutron-capture events are lower than the detection energy threshold.
\end{itemize}

These factors are affected by the concentration of gadolinium in the Gd-LS, which in turn can be determined from the time distribution of neutron captures as follows. The $^{241}$Am-Be neutron source was deployed to the center of the ND, and the neutrons emerging from the source were selected by the delayed coincidence technique described in Sec.~\ref{sec:calib}. The capture time of the neutrons of the selected sample was measured by fitting the temporal distribution of the delayed signals relative to the correlated prompt signals to an exponential function in the range of 20--200 $\mu$s. Simulations based on GEANT4~\cite{bib:agostinelli} were done to calculate the expected neutron-capture time for different gadolinium concentrations. The nominal composition of Gd-LS used in the simulations is tabulated in Table \ref{tab:gdls-composition}. The fraction of gadolinium by weight was changed from 0\% to 0.1\%, while the fractions for the other materials were adjusted accordingly. The dependence of the capture time on the concentration of gadolinium was then obtained from the simulated samples with the neutron source placed at the center of the ND. By parametrizing the dependence with a linear function, the best fit was obtained with $\chi^2/ndf = 17.6/14 = 1.25$ as
\begin{equation}
	\frac{1}{\tau_{nc}} = (0.2787 \pm 0.0026) \rho_{Gd} + (0.0049 \pm 0.0001) ,
	\label{eq:gd-conc-captime}
\end{equation}
where $\tau_{nc}$ is the neutron-capture time in microsecond and $\rho_{Gd}$ is the mass concentration of gadolinium in percentage. From the measured capture time, the mass concentration of gadolinium was deduced. It was found to be gradually decreasing from 0.069\% to 0.063\% starting from the middle of the experiment. The reason of the decrease was not clear but was thought to be related to the leakage of oxygen from air into the Gd-LS. Visual inspection of the interior of the ND at a later stage of the experiment revealed some white deposit accumulated at the bottom of the acrylic vessel. Further investigation on any possible relationship between the deposit and the decrease in gadolinium in the Gd-LS is needed. Nevertheless, this changing gadolinium concentration has been taken into account in the calculation of the neutron detection efficiency.

\begin{table}
	\centering
	\caption{The nominal composition of Gd-LS used in the GEANT4-based simulations.}
	\begin{ruledtabular}
		\begin{tabular}{ l c }
			Material name & Fractional mass \\
			\hline
			TS\_C\_of\_Graphite & 0.87740 \\
			TS\_H\_of\_Water & 0.12056 \\
			Oxygen & 0.00109 \\
			Gadolinium & 0.00063 \\
			Nitrogen & 0.00027 \\
			Sulfur & 0.00005 \\
		\end{tabular}
	\end{ruledtabular}
	\label{tab:gdls-composition}
\end{table}

\begin{figure*}
	\centering
	\includegraphics[width=\linewidth]{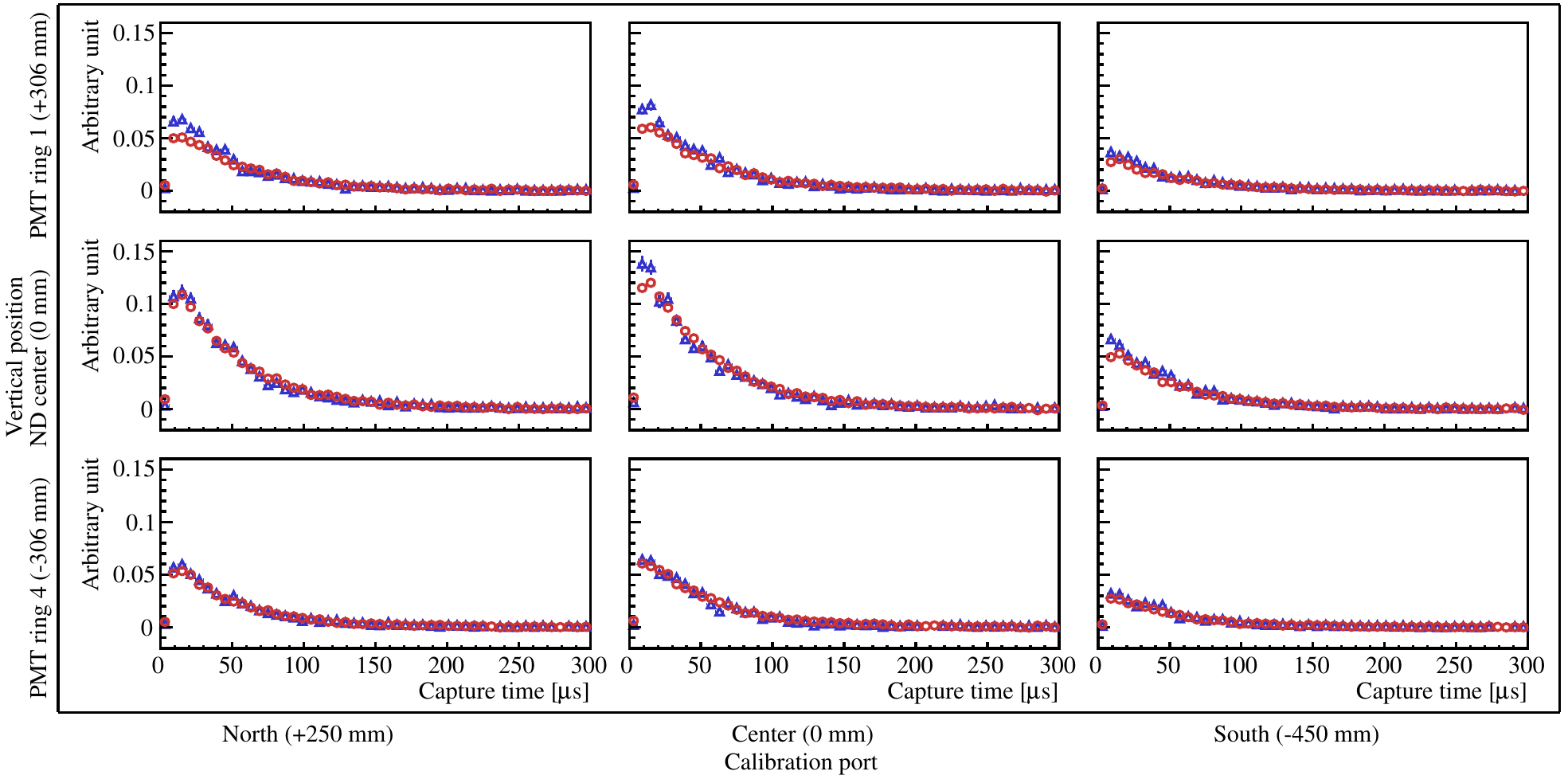}
	\caption{Comparison between the measured (circles) and the simulated (triangles) capture-time distributions of neutrons produced by the $^{241}$Am-Be source at different positions inside the ND. The drop in counts before 10 $\mu$s were due to thermalization of neutrons and the dead time of the DAQ system after the prompt signals.}
	\label{fig:calib-comparison-ambe-captime}
\end{figure*}

\begin{figure*}
	\centering
	\includegraphics[width=\linewidth]{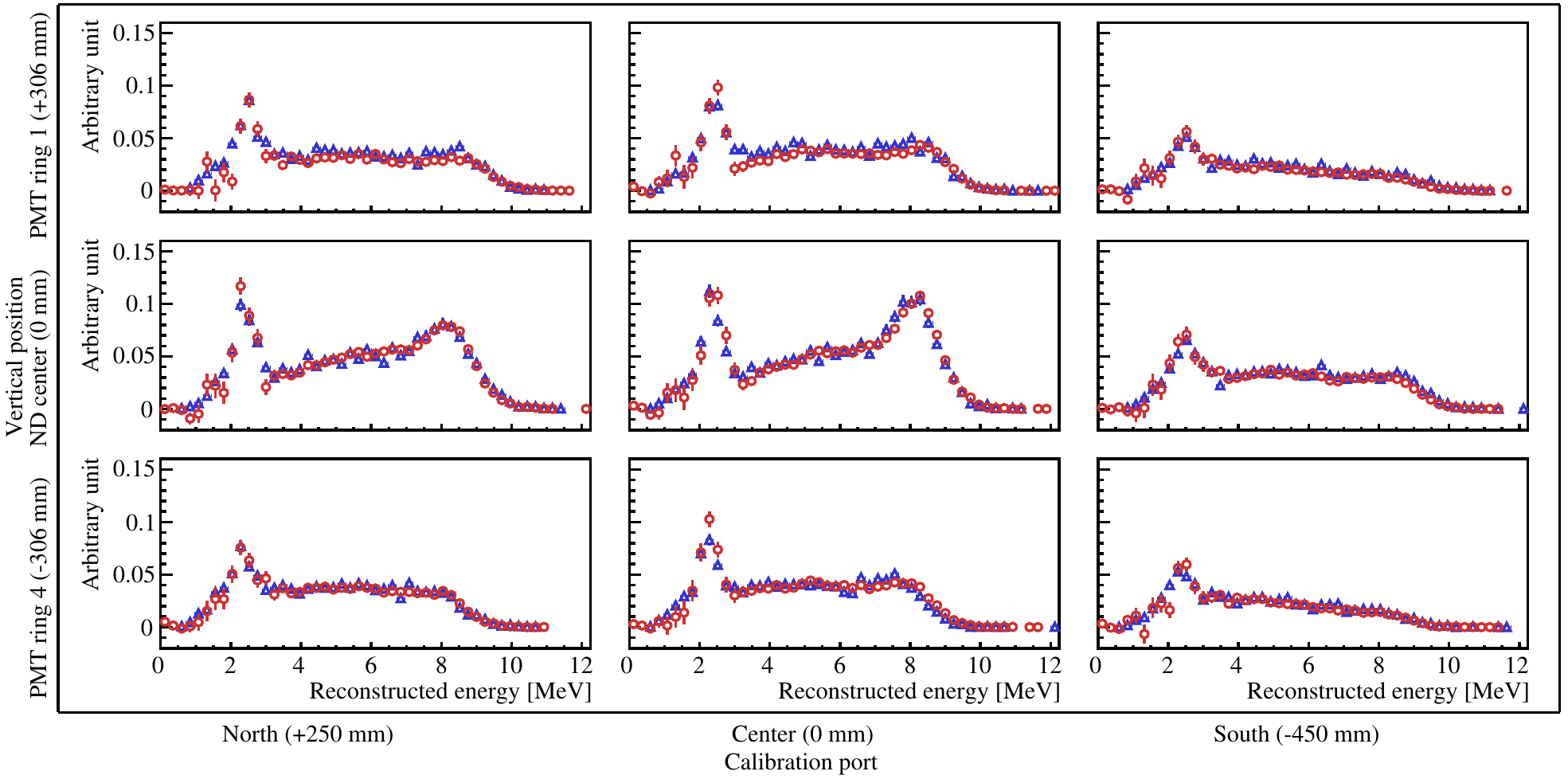}
	\caption{Comparison between the measured (circles) and the simulated (triangles) delayed energy distributions for neutrons emitted from the $^{241}$Am-Be source at different positions inside the ND. The peak at around 2.2 MeV was due to the capture of neutrons on hydrogen, while the broad peak at around 8 MeV was due to the gamma rays emitted after the capture of neutrons on gadolinium.}
	\label{fig:calib-comparison-ambe-energy}
\end{figure*}

The neutron detection efficiency $\varepsilon_{ND}$ was broken down into several constituting components as shown in Table \ref{tab:muin-eff-uncert}, with
\begin{equation}
	\varepsilon_{ND} = \varepsilon_{Gd} \varepsilon_{T} \varepsilon_{E} \varepsilon_{DAQ} \varepsilon_{Spill} .
	\label{eq:muin-eff-breakdown}
\end{equation}
Here, $\varepsilon_{Gd}$ is the ratio of the number of gadolinium-captured neutrons to the total number of neutron captures within the target. The quantity $\varepsilon_{T}$ is the efficiency of requiring the delayed signal to be less than 200 $\mu$s after the prompt signal. In the measurement, since we only counted neutrons that were captured on gadolinium with released energy greater than 4.6 MeV, $\varepsilon_{E}$ is the efficiency of this requirement. The central values of $\varepsilon_{Gd}$, $\varepsilon_{T}$, and $\varepsilon_{E}$ were evaluated with GEANT4-based simulations. The simulation code has been validated by comparing its results with experimental data as shown in Figs.~\ref{fig:calib-comparison-ambe-captime} and \ref{fig:calib-comparison-ambe-energy}. Differences between the measured and the simulated distributions were taken into account in estimating the systematic uncertainties. The uncertainty of $\varepsilon_{E}$ also included a 3\% uncertainty in the ND energy scale.

\begin{table}
	\centering
	\caption{Summary of absolute efficiencies and systematic uncertainties in the measurement of muon-induced neutrons.}
	\begin{ruledtabular}
		\begin{tabular}{ l c c }
			 & Efficiency & Uncertainty \\
			\hline
			Gd capture ratio, $\varepsilon_{Gd}$ & 0.800 & 0.010 \\
			Time cut, $\varepsilon_{T}$ & 0.866 & 0.008 \\
			Energy cut, $\varepsilon_{E}$ & 0.523 & 0.023 \\
			Live-time, $\varepsilon_{DAQ}$ & 0.889 & 0.021 \\
			Spilling, $\varepsilon_{Spill}$ & 0.891 & 0.148 \\
			\hline
			Overall, $\varepsilon_{ND}$ & 0.287 & 0.050 \\
		\end{tabular}
	\end{ruledtabular}
	\label{tab:muin-eff-uncert}
\end{table}

\begin{figure}
	\centering
	\includegraphics[width=\linewidth]{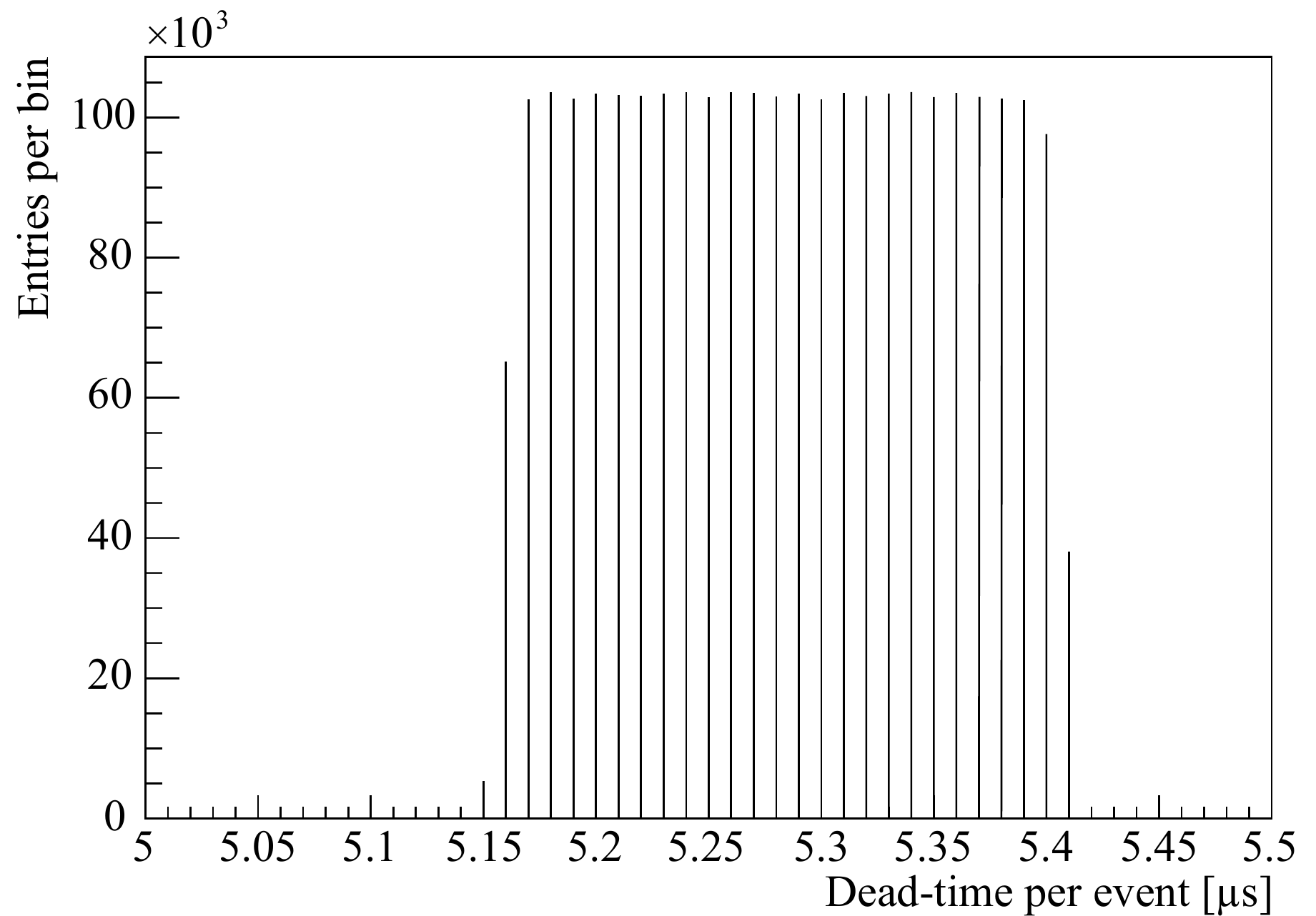}
	\caption{Measured dead-time distribution of the ND. The distribution was uniform between 5.17 and 5.40 $\mu$s.}
	\label{fig:nd-deadtime}
\end{figure}

The value of $\varepsilon_{DAQ}$ was evaluated with simulation. Triggers corresponding to neutron captures were generated according to the expected time distribution of the delayed signals. Triggers representing background events in a time window were sampled from a Poisson distribution. The fraction of background in the simulated sample was constrained to the measured amount seen in the data. Each simulated trigger was assigned with a dead time based on the measured dead-time distribution (Fig.~\ref{fig:nd-deadtime}). The percentage of triggers that did not fall into the dead time of any previous triggers (also known as unblocked triggers) was counted as a function of the number of neutrons generated in each time window. The result is listed in Table \ref{tab:daq-eff-breakdown}. The correction factor of the live time is a weighted average of the percentages, 
\begin{equation}
	\varepsilon_{DAQ} = \frac{\sum_{i=1}^{10} i \times f_{n,i} \times U_{i}}{\sum_{i=1}^{10} i \times f_{n,i}} ,
	\label{eq:daq-eff}
\end{equation}
where the summations run over different numbers of neutrons in a time window; $f_{n,i}$ is the frequency of observing $i$ neutrons in a time window, which is also listed in Table \ref{tab:daq-eff-breakdown}; and $U_{i}$ is the fraction of unblocked triggers when $i$ neutrons are present in a time window. The value of $\varepsilon_{DAQ}$ shown in Table \ref{tab:muin-eff-uncert} also includes the effect of a changing ND energy scale, where the influence was estimated by repeating the simulation with 3\% variation in the energy scale.

\begin{table}
	\centering
	\caption{Calculated fraction of unblocked triggers as a function of the number of neutrons in a time window and the measured frequency of the number of neutrons.}
	\begin{ruledtabular}
		\begin{tabular}{ l c c }
			Number of & Fraction of unblocked & Measured frequency, \\
			neutrons, $i$ & triggers, $U_{i}$ & $f_{n,i}$ \\
			\hline
			1 & 0.971 & $168.29 \pm 0.74$ \\
			2 & 0.918 & $31.64 \pm 2.10$ \\
			3 & 0.872 & $17.12 \pm 2.86$ \\
			4 & 0.833 & $10.82 \pm 1.73$ \\
			5 & 0.798 & $6.21 \pm 1.50$ \\
			6 & 0.767 & $3.51 \pm 0.99$ \\
			7 & 0.739 & $1.82 \pm 0.63$ \\
			8 & 0.714 & $1.05 \pm 0.50$ \\
			9 & 0.690 & $0.55 \pm 0.40$ \\
			10 & 0.669 & $0.36 \pm 0.31$ \\
			\hline
			Weighted &  &  \\
			average [Eq.~\eqref{eq:daq-eff}] & $0.893 \pm 0.014$ &  \\
		\end{tabular}
	\end{ruledtabular}
	\label{tab:daq-eff-breakdown}
\end{table}

Neutrons generated in the Gd-LS can drift out from the target, hence reducing the number of detectable neutrons. Similarly, some of the neutrons generated outside of the Gd-LS can drift in and be captured inside the target, increasing the number of detected neutrons. These two effects are termed as ``spill-out'' and ``spill-in,'' respectively. The net spilling fraction $\varepsilon_{Spill}$ was also evaluated with GEANT4-based simulation. We used the muon energies calculated with the MUSIC~\cite{bib:antonioli} code and the production model described in Ref.~\cite{bib:wang} to generate the muon-induced neutrons. Since the neutron energy spectrum is uncertain with a wide range of results reported in the literature \cite{bib:khalchukov, bib:perkins, bib:aglietta-lvd}, we have also generated neutrons with energy $E_{n}$ following the power law of the form $E_{n}^{-1}$ or $E_{n}^{-2}$ to cover all the possibilities. All the three models have a singularity at $E_{n} = 0$, and thus we have set the lowest neutron energy at 1 MeV. Neutrons with energy below 1 MeV were generated separately with an uniform distribution. The net spilling fraction evaluated from the first model was very close to the mean value of the other two models. The central value of $\varepsilon_{Spill}$ was the average result of the three models, and its uncertainty covered the span of the three results.

\section{Results and Discussion\label{sec:result}}

\subsection{Muon flux\label{sec:result-muon}}

\begin{figure}
	\centering
	\includegraphics[width=\linewidth]{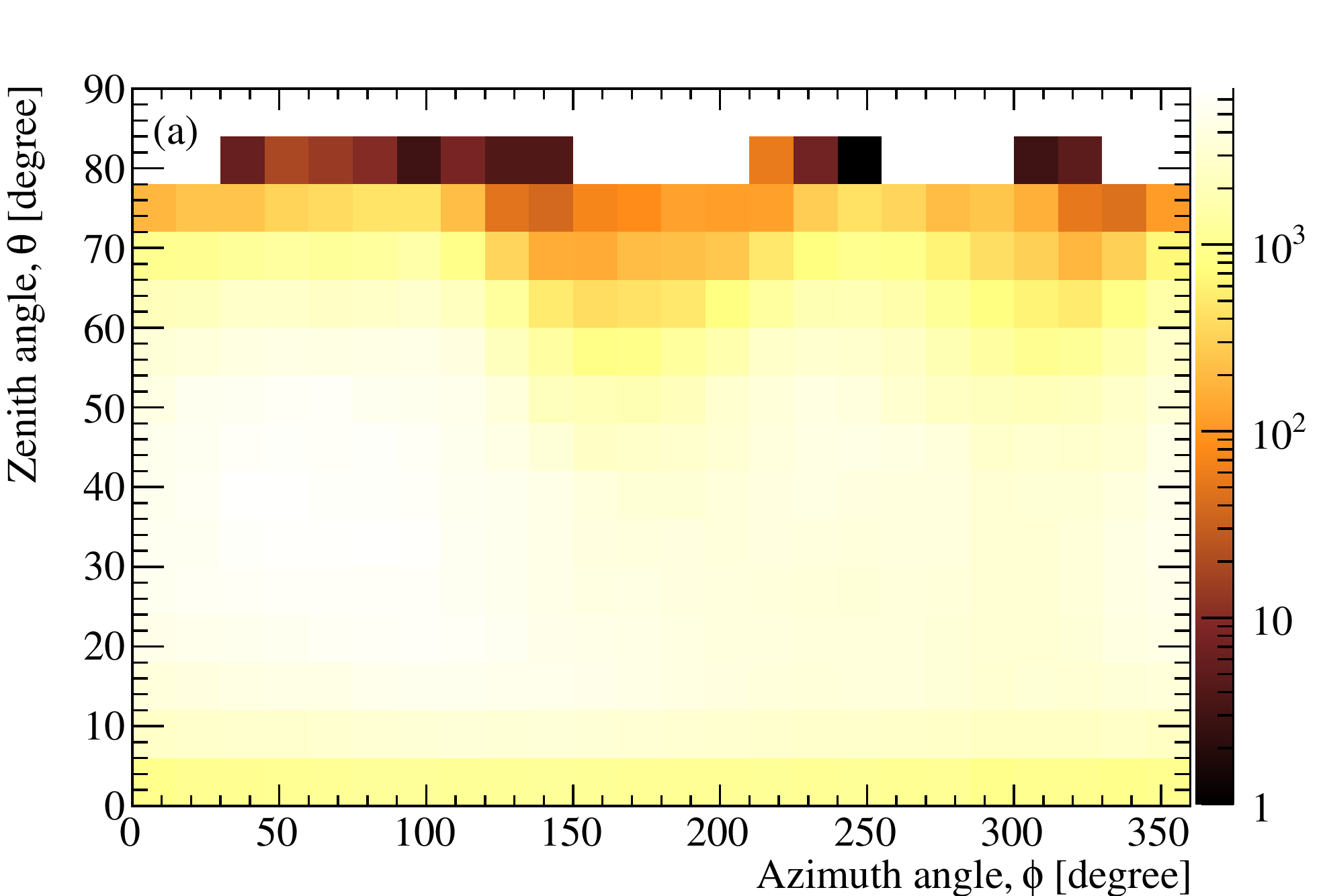}
	\includegraphics[width=\linewidth]{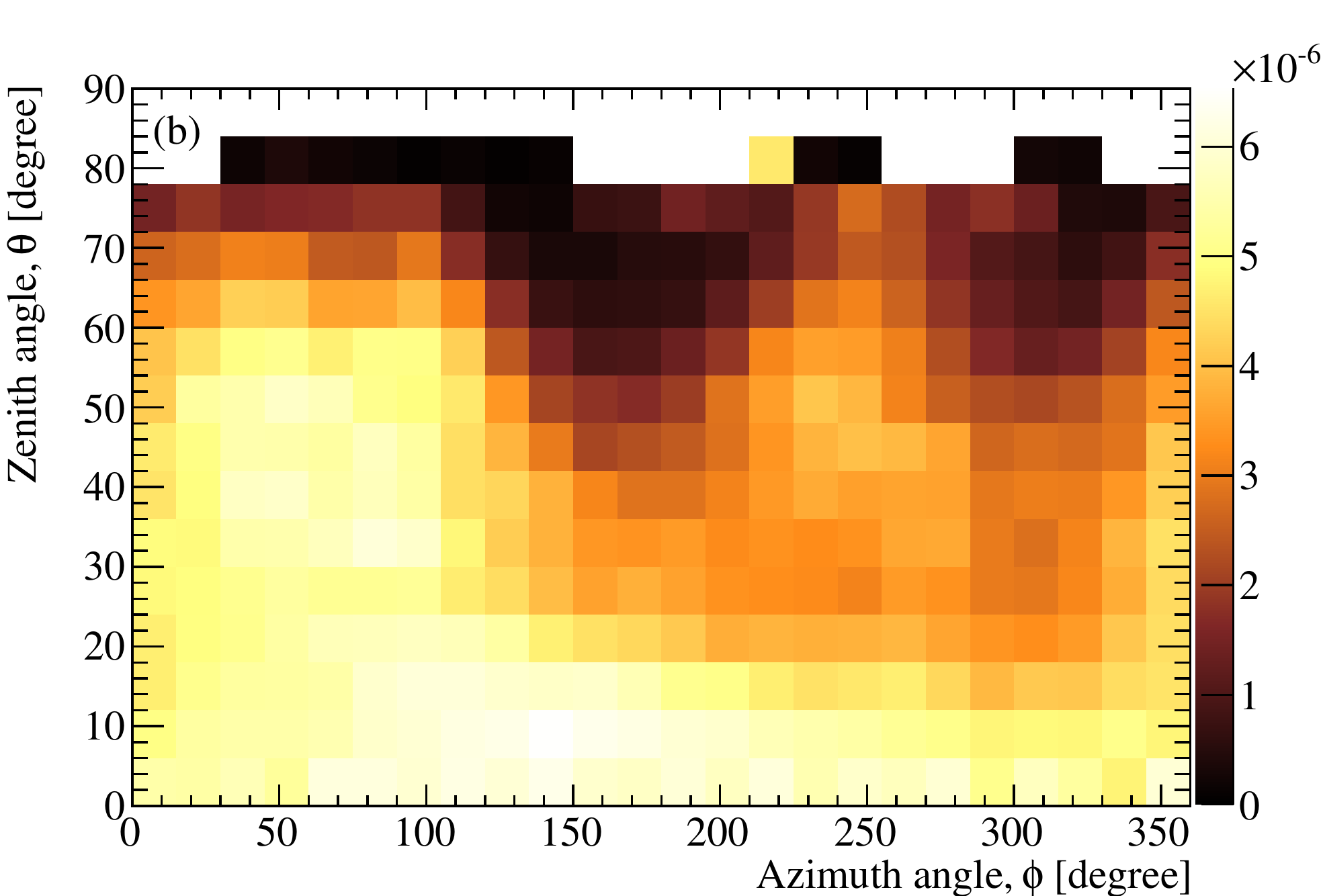}
	\caption{(a) Measured number of muons and (b) muon intensity in the unit of cm$^{-2}$s$^{-1}$sr$^{-1}$. Both figures were drawn with a bin size of $6^{\circ} \times 15^{\circ}$. The statistic uncertainty for each bin was in general better than 10\%, except for $\theta > 72^{\circ}$ where some bins had less than 100 events. The systematic uncertainty was about 10\%.}
	\label{fig:mu-intensity}
\end{figure}

\begin{figure}
	\centering
	\includegraphics[width=\linewidth]{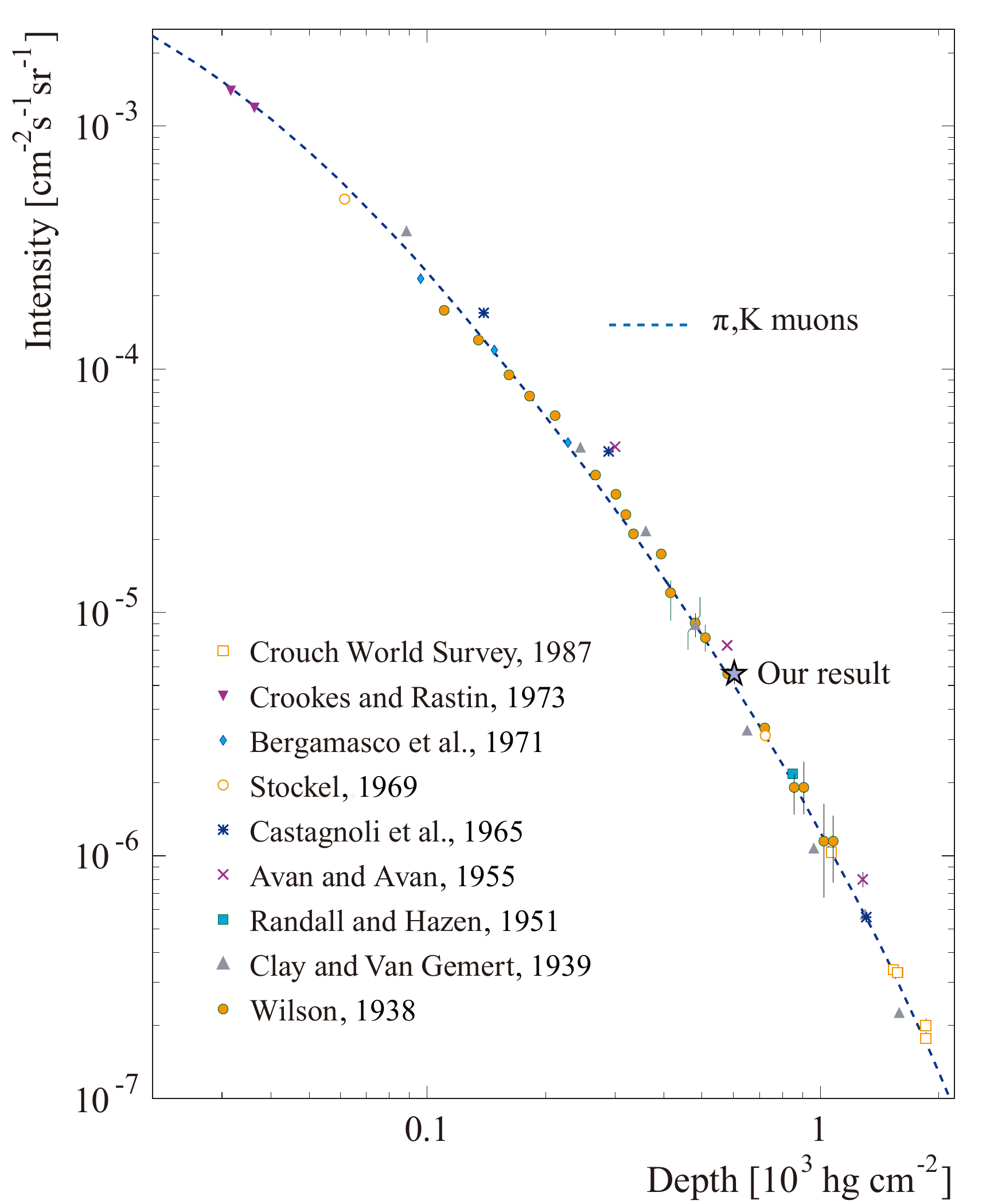}
	\caption{Measured vertical muon flux as a function of depth, adopted from Ref.~\cite{bib:bugaev}. The current result is shown as a star, with error bars smaller than the size of the symbol.}
	\label{fig:mu-intensity-literature}
\end{figure}

The double differential muon intensity as a function of the zenith angle $\theta$ and azimuth angle $\phi$, $I_{\mu}(\theta,\phi)$, is given by
\begin{equation}
	I_{\mu}(\theta,\phi) = \frac{N_{\mu}(\theta,\phi)}{A_{MT}(\theta,\phi) \varOmega(\theta,\phi) \tau_{MT} \varepsilon_{MT}(\theta,\phi) \varepsilon_{Fit}} ,
	\label{eq:mu-intensity}
\end{equation}
where $N_{\mu}$ is the number of selected muon events, $A_{MT}$ is the projected area of the MT in the horizontal plane, $\varOmega$ is the solid angle subtended by the bin, $\tau_{MT}$ is the live time of the measurement, and $\varepsilon_{MT}$ ($\varepsilon_{Fit}$) is the MT (fitness requirement) efficiency described in Sec.~\ref{sec:eff-mt}. Based on $9.5 \times 10^{5}$ reconstructed and selected muon events, the measured $N_{\mu}(\theta,\phi)$ and $I_{\mu}(\theta,\phi)$ are shown in Fig.~\ref{fig:mu-intensity}. The average vertical muon intensity defined by
\begin{equation}
	\langle I_{\mu} \rangle(\theta < 10^{\circ}) = \frac{1}{N_{bin}^{\theta < 10^{\circ}}} \sum_{\theta = 0^{\circ}}^{\theta < 10^{\circ}} \sum_{\phi = 0^{\circ}}^{\phi < 360^{\circ}} I_{\mu}(\theta,\phi) ,
	\label{eq:mu-intensity-vertical}
\end{equation}
where $N_{bin}^{\theta < 10^{\circ}} = 10 \times 360 = 3600$ is the number of bins in the summations, was measured to be $\langle I_{\mu} \rangle(\theta < 10^{\circ}) = (5.7 \pm 0.6) \times 10^{-6}$ cm$^{-2}$s$^{-1}$sr$^{-1}$. As shown in Fig.~\ref{fig:mu-intensity-literature}, the current result is in good agreement with the interpolation to our overburden from the other measurements compiled in Ref.~\cite{bib:bugaev}.

\begin{figure}
	\centering
	\includegraphics[width=\linewidth]{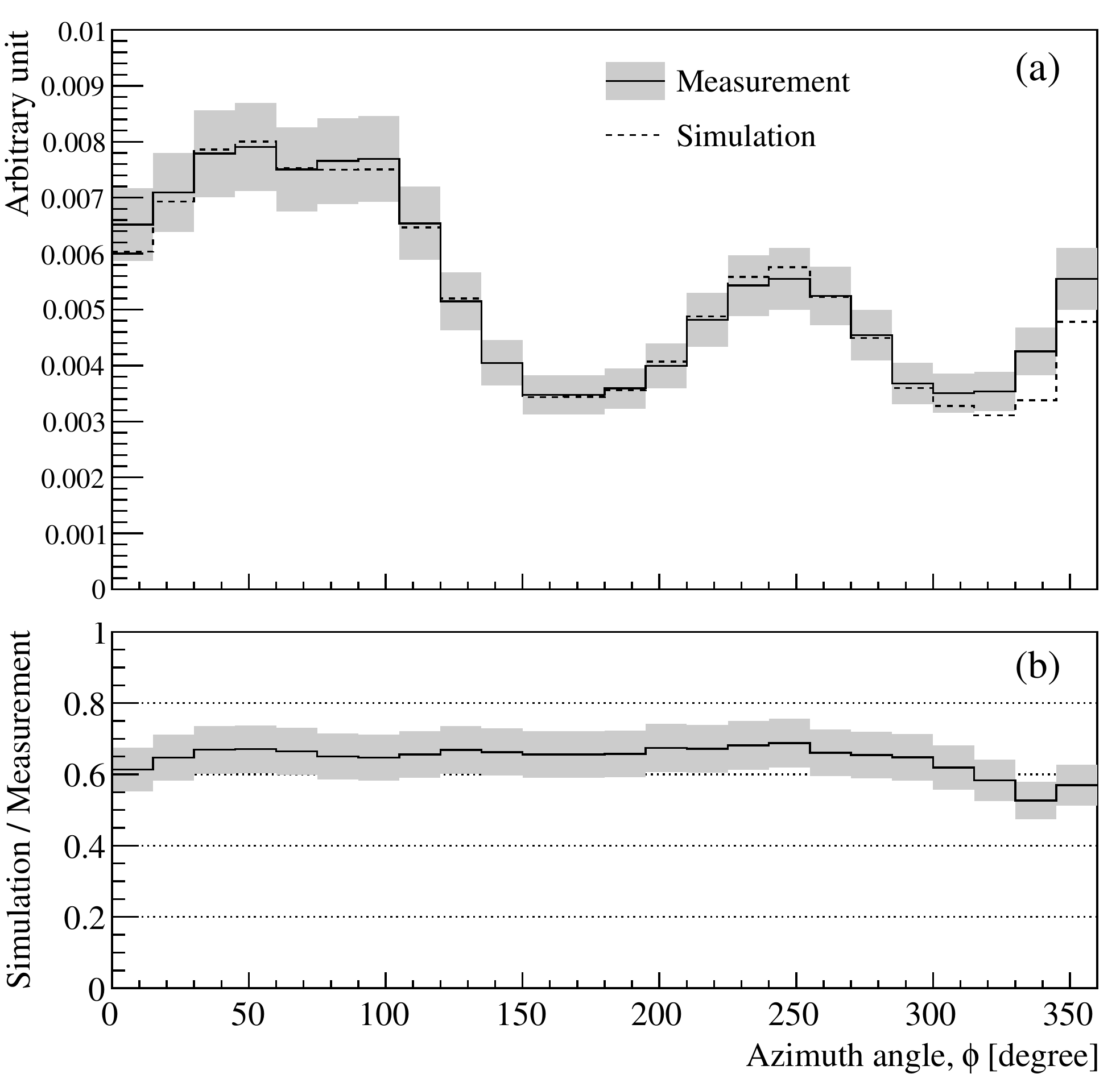}
	\caption{Comparison between the measured and the simulated $\langle F_{\mu}(\phi) \rangle$ distributions. (a) Both curves are normalized by their respective total areas between 90$^{\circ}$ and 270$^{\circ}$ in azimuth. The error band is shown in shade for the measured distribution. (b) Ratio between the absolute fluxes.}
	\label{fig:mu-flux-comparison-phi}
\end{figure}

\begin{figure}
	\centering
	\includegraphics[width=\linewidth]{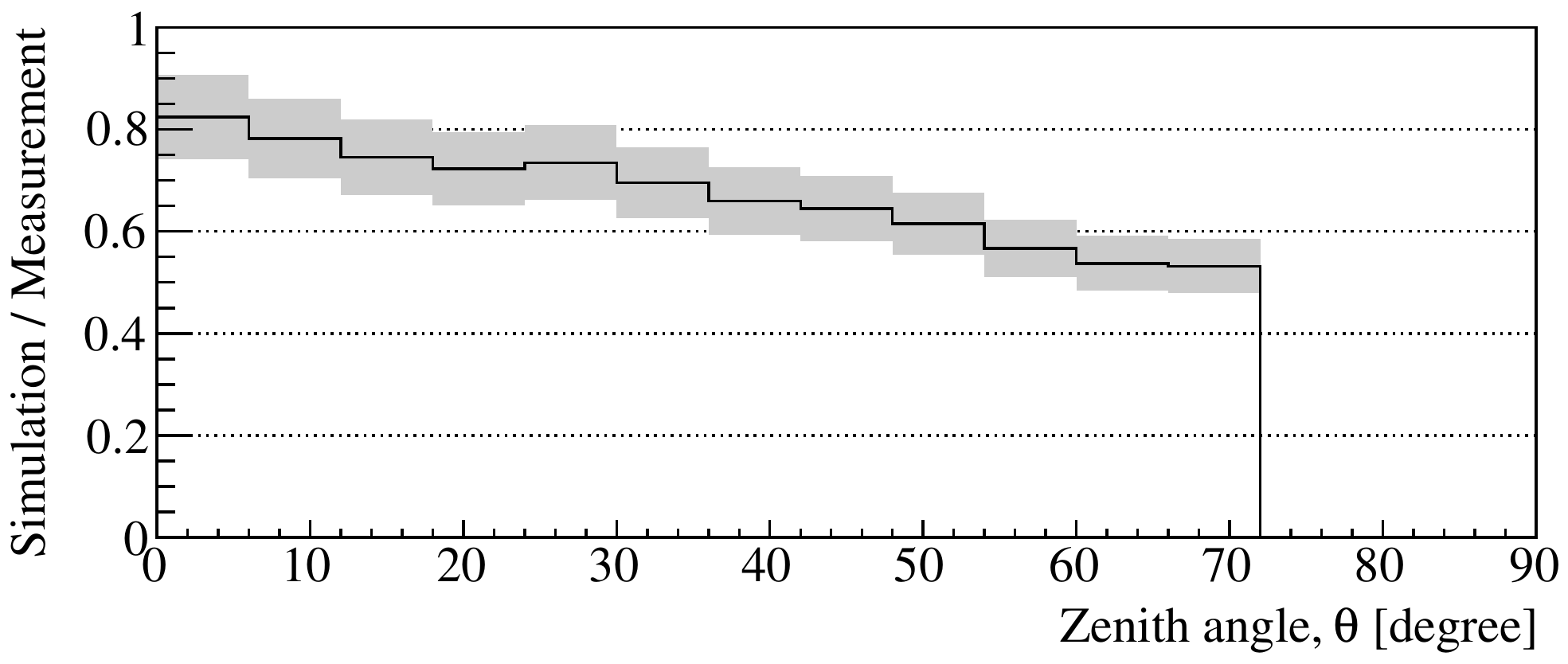}
	\caption{Ratio between the measured and the simulated $\langle F_{\mu}(\theta) \rangle$ distributions. The error band is shown in shade. The plot is limited to 72$^{\circ}$ due to the coverage of the topographical map used in the simulation.}
	\label{fig:mu-flux-comparison-theta}
\end{figure}

Other forms of differential muon flux that can be extracted from the data are
\begin{equation}
	\langle F_{\mu}(\phi) \rangle = \sum_{\theta = 0^{\circ}}^{\theta < 90^{\circ}} I_{\mu}(\theta,\phi) \varOmega(\theta,\phi)
	\label{eq:mu-flux-phi}
\end{equation}
and
\begin{equation}
	\langle F_{\mu}(\theta) \rangle = \sum_{\phi = 0^{\circ}}^{\phi < 360^{\circ}} I_{\mu}(\theta,\phi) \varOmega(\theta,\phi) ,
	\label{eq:mu-flux-theta}
\end{equation}
which can be compared to the prediction obtained from the modified Gaisser parametrization for describing the muon distribution at the surface of the Earth \cite{bib:gaisser, bib:guan}: 
\begin{align}
	\frac{dF_{\mu}}{dE_{\mu} d\varOmega} &\approx \frac{0.14}{\rm{cm^{2} \ sr \ s \ GeV}} \left[ \frac{E_{\mu}}{\rm{GeV}} \left( 1 + \frac{3.64 \ \rm{GeV}}{E_{\mu} (\cos \theta^{*})^{1.29}} \right) \right]^{-2.7} \notag \\ &\times \left( \frac{1}{1 + \frac{1.1 E_{\mu} \cos \theta^{*}}{115 \ \rm{GeV}}} + \frac{0.054}{1 + \frac{1.1 E_{\mu} \cos \theta^{*}}{850 \ \rm{GeV}} } \right) ,
	\label{eq:mod-gaisser}
\end{align}
where $F_{\mu}$ is the muon flux, $E_{\mu}$ is the muon energy, and 
\begin{equation}
	\cos \theta^{*} = \sqrt{\frac{(\cos \theta)^{2} + P_{1}^{2} + P_{2}(\cos \theta)^{P_{3}} + P_{4}(\cos \theta)^{P_{5}}}{1 + P_{1}^{2} + P_{2} + P_{4}}} ,
	\label{eq:cos-theta-star}
\end{equation}
with the parameters ($P_{1}$ = 0.102573, $P_{2}$ = -0.068287, $P_{3}$ = 0.958633, $P_{4}$ =  0.0407253, $P_{5}$ = 0.817285) given in Ref.~\cite{bib:chirkin}. Assuming a rock density of 2.60 g$\cdot$cm$^{-3}$, the muons simulated according to Eq.~\eqref{eq:mod-gaisser} were then transported through the overburden using the MUSIC code and a digitized three-dimensional topographical map of the experimental site for predicting the underground muon flux and energy. The topographical map was large enough to cover the zenith angles up to about 72$^{\circ}$. The simulation predicted the average muon energy inside the laboratory to be about 120 GeV and the integrated muon flux to be approximately $1 \times 10^{-5}$ cm$^{-2}$s$^{-1}$. The simulation result is compared to the measured $\langle F_{\mu}(\phi) \rangle$ distribution in Fig.~\ref{fig:mu-flux-comparison-phi}. The general profile of both distributions are similar. Some inconsistencies can be seen around $\phi$ = 340$^{\circ}$, which may be due to the lack of a detailed local geological map for the simulation. However, the simulation underestimated the overall muon flux by about 35\%, which cannot be explained by an incorrect rock density being used in the simulation (using a rock density of 2.50 g$\cdot$cm$^{-3}$ reduces the deficit to about 25\%) or the uncertainty in the measurement (about 10\%). The deficit in the simulation showed a dependence on the zenith angle as shown in Fig.~\ref{fig:mu-flux-comparison-theta}. The dependence was quite linear with a dropping rate of about 0.44\% per degree from 0$^{\circ}$ to 72$^{\circ}$. Neither the MT efficiency nor the fitness requirement efficiency showed this dependence on the zenith angle. Changing the rock density in the simulation could not resolve the dependence, unless the density was reduced to 2.30 g$\cdot$cm$^{-3}$ in the low-altitude regions of the mountains. The source of the dependence could also be due to the description of surface muon distribution in Eq.~\eqref{eq:mod-gaisser} or the transport of muons by the MUSIC code. However, because of the lack of a detailed local geological map, we cannot come to a definite conclusion.

\subsection{Muon-induced neutrons\label{sec:result-muin}}

\begin{figure}
	\centering
	\includegraphics[width=\linewidth]{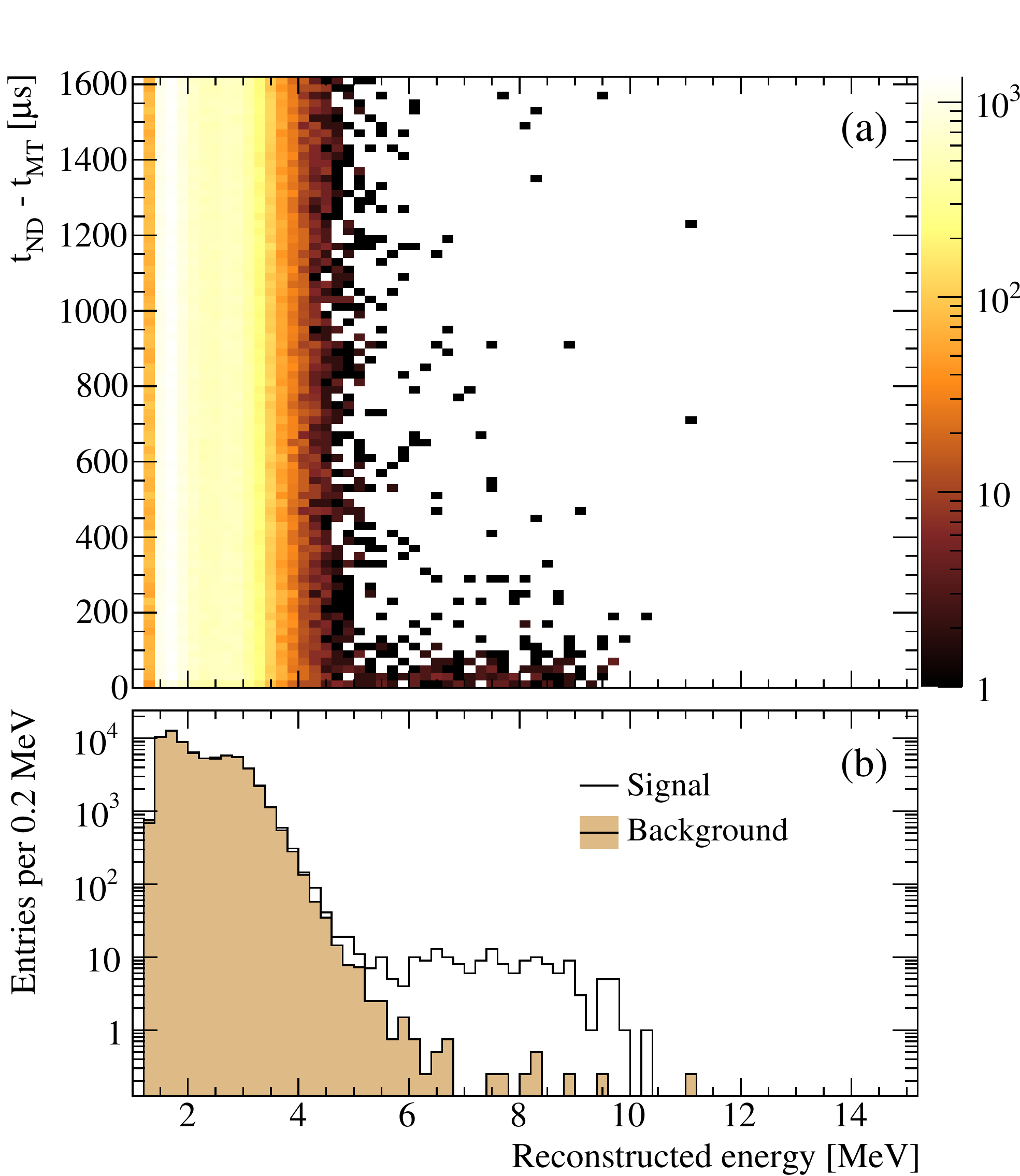}
	\caption{(a) Energy and time distribution of ND signals following prompt MT signals. Muon-induced neutron-capture events can be seen around 8 MeV from 0 to 200 $\mu$s. (b) Comparison of ND energy spectra in the signal time window (10--210 $\mu$s) and the background time window (800--1600 $\mu$s). The background spectrum was scaled by a factor of 0.25.}
	\label{fig:muin-energy-time}
\end{figure}

\begin{figure}
	\centering
	\includegraphics[width=\linewidth]{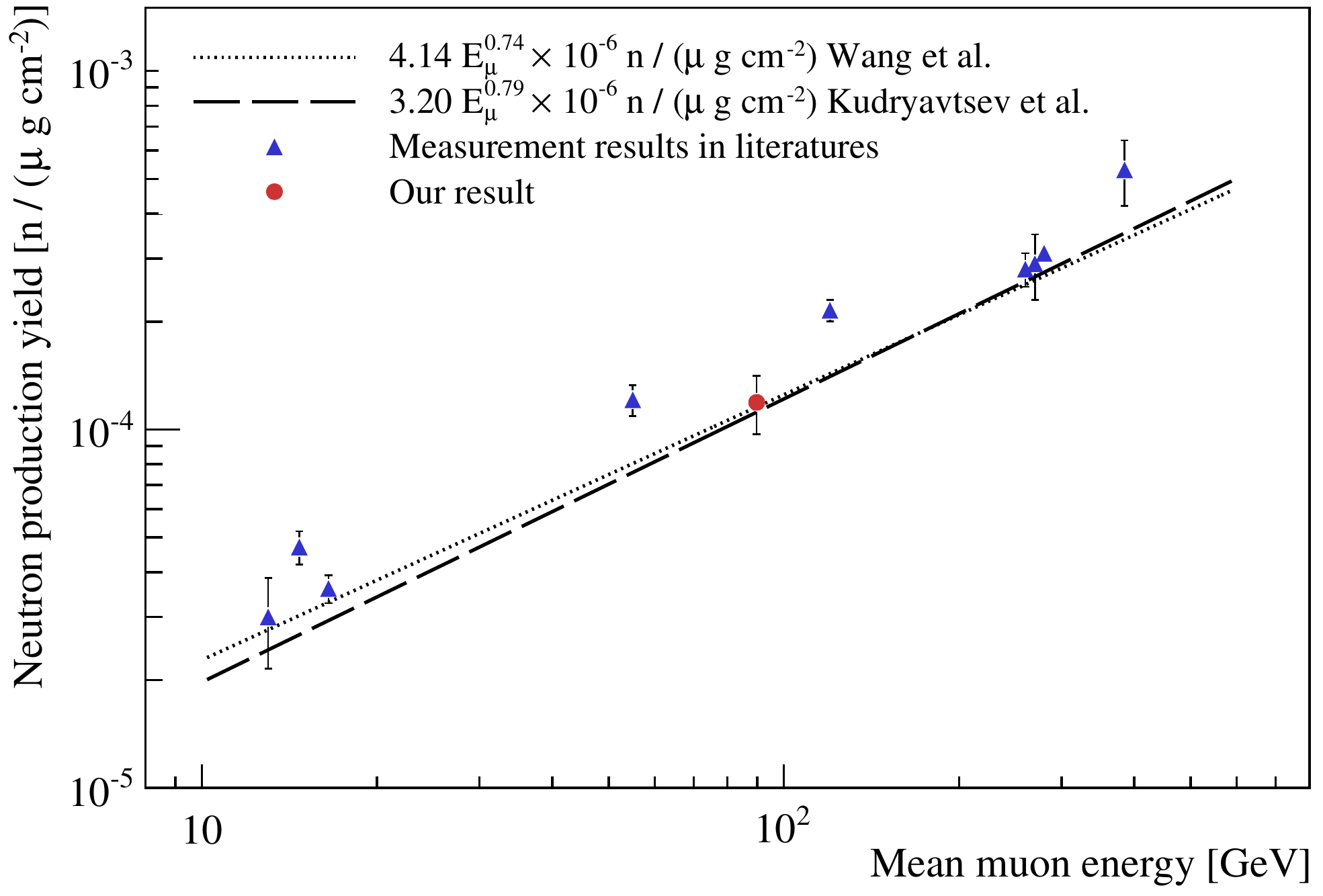}
	\caption{Total neutron production yield as a function of muon energy. The dot represents our measurement result. The lines show the fitting result of FLUKA simulation studies \cite{bib:wang, bib:kudryavtsev}. The triangles, from left to right, are other measurement results from, respectively, the Cosmic-ray Underground Background Experiment (CUBE) in the Stanford Underground Facility \cite{bib:hertenberger} (depth 20 m.w.e.), a gypsum mine \cite{bib:bezrukov} (depth 25 m.w.e.), the Palo Verde experiment \cite{bib:boehm} (depth 32 m.w.e.), a salt mine \cite{bib:bezrukov} (depth 316 m.w.e), the Artemovsk Scientific Station \cite{bib:enikeev} (depth 570 m.w.e.), the KamLAND experiment \cite{bib:abe} (depth 2700 m.w.e.), the Large Volume Detector (LVD) in Gran Sasso \cite{bib:persiani} (depth 3650 m.w.e.), the Borexino experiment \cite{bib:bellini} (depth 3800 m.w.e.), and the Liquid Scintillation Detector (LSD) in Mont Blanc \cite{bib:aglietta-lsd} (depth 5200 m.w.e.). For the measurement results, the abscissa corresponds to the average muon energy at the experimental depth.}
	\label{fig:muin-comparison}
\end{figure}

In this part of the analysis, the reconstructed muon track was required to be inside the Gd-LS, corresponding to having a maximum zenith angle of about 35$^{\circ}$. This requirement, together with the strict selection criteria imposed on the prompt signals as described in Sec.~\ref{sec:select}, reduced the number of detected muons by an order of magnitude to 93,061. Their mean energy in the underground laboratory was calculated with the MUSIC to be $(89.8 \pm 2.9)$ GeV, where the uncertainty was estimated by varying the rock density and the altitude of the laboratory by $\pm 0.1$ g$\cdot$cm$^{-3}$ and $\pm 10$ m, respectively. Figure \ref{fig:muin-energy-time} shows the distribution of ND signals following the prompt signals. The total number of events in the signal (background) window after the energy requirement of $\geq$ 4.6 MeV was 225 (164). Since the time window for determining the background was four times wider than the signal one, the number of neutron candidates was $225 - 164/4 = 184 \pm 15$. The neutron yield $Y_{n}$ is calculated as
\begin{equation}
	Y_{n} = \frac{N_{n}}{N_{\mu} L_{\mu} A_{ND} \rho_{LS} \varepsilon_{ND}} ,
	\label{eq:muin-yield}
\end{equation}
where $N_{n}$ is the number of neutron candidates, $N_{\mu}$ is the number of muons, $L_{\mu}$ is the mean path length of the incident muons in the Gd-LS, $A_{ND}$ is the acceptance of the ND, $\rho_{LS}$ is the density of the Gd-LS, and $\varepsilon_{ND}$ is the detection efficiency of neutrons described in Sec.~\ref{sec:eff-nd}. The muon path lengths were calculated from the intercepts between the reconstructed muon tracks and the boundary of the Gd-LS volume. The value of $L_{\mu}$ was determined to be $(81.6 \pm 1.0)$ cm. The value of $A_{ND}$ was evaluated with Monte Carlo simulation. Muons were generated according to the underground distributions produced with the MUSIC package as described in Sec.~\ref{sec:result-muon} and were selected with the prompt-signal criteria. Neutrons were generated at some radial distances along each muon track, following a distribution \cite{bib:hagner}, 
\begin{equation}
	\Phi(r) = \frac{1}{166}e^{-0.0403 r^{2}} + \frac{1}{1223}e^{-0.098 r} ,
	\label{eq:muin-lateral-dist}
\end{equation}
where $r$ is the perpendicular distance from the muon track in centimeters. The distribution $\Phi(r)$ is normalized as $\int_{0}^{\infty} \Phi(r) 2\pi r dr = 1$. The uncertainty of Eq.~\eqref{eq:muin-lateral-dist} is about 3\% for $r$ smaller than about 10 cm and increases to about 18\% for $r \approx 30$ cm \cite{bib:hagner-phd}. The acceptance expressed as the fraction of neutrons generated within the target volume was determined to be $(82.7 \pm 3.7)$\%. Given that the density $\rho_{LS} = (0.855 \pm 0.004)$ g$\cdot$cm$^{-3}$, the muon-induced neutron yield was determined to be $Y_{n} = (1.19 \pm 0.08 \textnormal{(stat)} \pm 0.21 \textnormal{(syst)}) \times 10^{-4}$ neutrons/($\mu\cdot$g$\cdot$cm$^{-2}$). A comparison of our measured neutron yield at $\left\langle E_{\mu} \right\rangle = (89.8 \pm 2.9)$ GeV with other measurements and simulation results using FLUKA~\cite{bib:ferrari} in the literature is shown in Fig.~\ref{fig:muin-comparison}. Our result and the other recent measurements \cite{bib:hertenberger, bib:boehm, bib:abe, bib:persiani, bib:bellini} are about 10\% higher than the expectations obtained with FLUKA. For the results reported before 1990 \cite{bib:bezrukov, bib:enikeev, bib:aglietta-lsd}, the neutron yields are 30\% to 40\% higher than the predictions. The disagreement between the experimental results and predictions is insensitive to the energy distribution of the underground muons used in the simulation \cite{bib:kudryavtsev}. It has been suggested that the problem might be related to the lower $^{11}$C production rate in FLUKA \cite{bib:bellini}. The systematic difference between the pre-1990 and post-1990 experimental results is unknown, but could be due to different treatments of correlated neutron background generated in rock and detector materials. The dependence of the measured neutron yields on the mean muon energy can be modeled with a power law: 
\begin{equation}
	Y_{n} = a \times 10^{-6} \left\langle E_{\mu} \right\rangle^b .
	\label{eq:muin-power-law}
\end{equation}
The best fits yielded $a = 6.52 \pm 1.54$ and $b = 0.73 \pm 0.05$ for pre-1990 data, $a = 4.23 \pm 0.72$ and $b = 0.76 \pm 0.03$ for post-1990 data, and $a = 5.69 \pm 0.68$ and $b = 0.71 \pm 0.02$ for all measurements. In any case, the energy dependence of the simulated neutron yields is in reasonably good agreement with the measurements.

\section{Conclusions\label{sec:conclude}}

The Aberdeen Tunnel experiment has measured the muon flux and the muon-induced neutron yield at a moderate depth of 611 m.w.e. The vertical muon intensity was measured to be $I_{\mu} = (5.7 \pm 0.6) \times 10^{-6}$ cm$^{-2}$s$^{-1}$sr$^{-1}$. We found good agreement with the other measurements. However, simulation done with MUSIC using the modified Gaisser parametrization of the surface muon distribution [Eq.~\eqref{eq:mod-gaisser}] underestimates the muon flux by 20\% to 30\%. We have also obtained a muon-induced neutron yield $Y_{n} = (1.19 \pm 0.08 \textnormal{(stat)} \pm 0.21 \textnormal{(syst)}) \times 10^{-4}$ neutrons/($\mu\cdot$g$\cdot$cm$^{-2}$) for the linear-alkyl-benzene-based liquid scintillator. A fit to the recent results with scintillator targets at different depths gave the neutron yield $Y_{n} = (4.23 \pm 0.72) \times 10^{-6} \left\langle E_{\mu} \right\rangle^{0.76 \pm 0.03}$ neutrons/($\mu\cdot$g$\cdot$cm$^{-2}$), which is in reasonable agreement with the predictions derived from FLUKA-based simulation.


\begin{acknowledgments}

This work is partially supported by grants from the Research Grant Council of Hong Kong Special Administrative Region, China (Projects No. HKU703307P, No. HKU704007P, No. CUHK 1/07C, and No. CUHK3/CRF/10); University Development Fund and Small Project Funding of The University of Hong Kong; Vice-Chancellor's One-off Discretionary Fund of The Chinese University of Hong Kong; the Office of Nuclear Physics, Office of High Energy Physics, Office of Science, US Department of Energy, under Contracts No. DE-AC-02-05CH11231 and No. DE-AC-02-98CH10886; and the National Science Council in Taiwan and MOE program for Research of Excellence at National Taiwan University and National Chiao Tung University. The authors would like to thank the Transport Department, The Government of the Hong Kong Special Administrative Region, for providing the underground facilities and Serco Group, plc, for their cooperation and support in the Aberdeen Tunnel.

\end{acknowledgments}

\bibliography{AbT-PRD}

\end{document}